\documentclass[11pt,a4paper]{article}
\pdfoutput=1

\newif\ifpublic\publictrue

\ifpublic\else\usepackage{showkeys}\fi
\usepackage[bookmarks=true,bookmarksnumbered,hyperfigures=true,hidelinks]{hyperref}
\usepackage[nosort]{cite}
\usepackage[parsep]{collref}
\usepackage[english]{babel}
\usepackage[T1]{fontenc}
\usepackage[latin1]{inputenc}
\usepackage{amsfonts,amsbsy,dsfont,euscript,mathrsfs,fixmath}
\usepackage{amsmath,amssymb}
\usepackage{enumerate}
\usepackage{array}
\usepackage{delimset}

\def\showkeysrefformat#1{{\normalfont\tiny\ttfamily#1}}
\makeatletter
\def\SK@@ref#1>#2\SK@{{\@inlabelfalse\leavevmode\vbox to\z@{\vss\SK@refcolor\rlap{\vrule\raise .75em \hbox{\showkeysrefformat{#2}}}}}}
\makeatother

\usepackage[a4paper,text={160mm,247mm},centering]{geometry}
\allowdisplaybreaks[3]
\numberwithin{equation}{section}
\def\[{\begin{equation}\begin{aligned}}
\def\]{\end{aligned}\end{equation}}
\newcommand{\nn}{\nonumber}
\newcommand{\nln}{\nonumber\\}

\usepackage[font=small,labelfont=bf,width=0.85\textwidth]{caption}

\expandafter\def\expandafter\bfseries\expandafter{\bfseries\ifmmode\else\boldmath\fi}
\expandafter\def\expandafter\mdseries\expandafter{\mdseries\ifmmode\else\unboldmath\fi}
\expandafter\def\expandafter\normalfont\expandafter{\normalfont\ifmmode\else\unboldmath\fi}

\RequirePackage{verbatim}

\makeatletter
\newwrite\bibinl@out
\newenvironment{bibtex}[1][\jobname]{%
\immediate\openout\bibinl@out #1.bib%
\immediate\write\bibinl@out{\@percentchar generated from `\jobname' starting line \the\inputlineno^^J}%
\def\verbatim@processline{\immediate\write\bibinl@out{\the\verbatim@line}}%
\@bsphack\let\do\@makeother\dospecials\catcode`\^^M\active\verbatim@start%
}
{\immediate\closeout\bibinl@out\@esphack}
\makeatother

\let\barefrac=\frac
\renewcommand{\frac}[2]{\mathinner{\barefrac{#1}{#2}}}

\let\baresqrt=\sqrt
\makeatletter
\renewcommand{\sqrt}{\@ifnextchar[\@sqrt@space@a\@sqrt@space@b}
\def\@sqrt@space@a[#1]#2{\mathinner{\mathchoice{\mkern-3mu}{\mkern-3mu}{}{}\baresqrt[#1]{#2}}}
\def\@sqrt@space@b#1{\mathinner{\mathchoice{\mkern-3mu}{\mkern-3mu}{}{}\baresqrt{#1}}}
\makeatother

\makeatletter
\let\per@dot@old=\.
\def\.{\ifmmode\def\per@dot@sel{\mkern3mu}\else\def\per@dot@sel{\per@dot@old}\fi\per@dot@sel}
\makeatother

\let\barefootnote=\footnote
\renewcommand{\footnote}[1]{\barefootnote{#1\vspace{3pt}}}

\newcommand{\vfrac}[2]{\ifmmode\mathinner{\textstyle^{#1}\!/\!_{#2}}\else$^{#1}\!/\!_{#2}$\fi}

\newcommand{\identity}{\mathds{1}}

\DeclareMathOperator{\diag}{diag}

\DeclareMathOperator{\STr}{STr}

\DeclareMathOperator{\sdet}{sdet}

\newcommand{\Complex}{\mathds{C}}
\newcommand{\Integer}{\mathds{Z}}

\DeclareMathOperator{\arsinh}{arsinh}

\newcommand{\ind}[1]{{\scriptscriptstyle{#1}}}

\newcommand{\alg}[1]{\mathfrak{#1}}
\newcommand{\grp}[1]{\mathrm{#1}}

\DeclareMathOperator{\Ad}{Ad}

\newcommand{\com}[2]{[#1,#2]}
\newcommand{\anticom}[2]{\{#1,#2\}}
\newcommand{\mcom}[2]{[#1,#2\}}

\def\<{\big\langle}
\def\>{\big\rangle}

\newcommand{\geom}[1]{\mathrm{#1}}

\newcommand{\AdS}{\geom{AdS}}

\newcommand{\Sp}{\geom{S}}

\newcommand{\To}{\geom{T}}

\newcommand{\extder}{\mathrm{d}}
\newcommand{\intder}{\iota}

\newcommand{\Act}{S}

\def\wasyfamily{\fontencoding{U}\fontfamily{wasy}\selectfont}
\def\Circle{\mbox{\wasyfamily\char35}}
\newcommand\RRF{\mathcal{F}}

\newcommand\bidef{\bar{\eta}}
\def\kp{\kappa_+}
\def\km{\kappa_-}
\usepackage{graphicx}
\usepackage{xhfill}
\usepackage{hyphenat}

\providecommand{\href}[2]{#2}

\makeatletter
\def\mr@ignsp#1 {\ifx\:#1\@empty\else #1\expandafter\mr@ignsp\fi}
\newcommand{\multiref}[1]{\begingroup%
	\xdef\mr@no@sparg{\expandafter\mr@ignsp#1 \: }%
	\def\mr@comma{}\def\mr@dash{-}%
	\@for\mr@refs:=\mr@no@sparg\do{%
		\ifx\mr@refs\mr@dash\def\mr@comma{}--\else%
		\mr@comma\def\mr@comma{,}\ref{\mr@refs}\fi}%
	\endgroup}
\renewcommand{\eqref}[1]{(\multiref{#1})}
\makeatother

\makeatletter
\newcommand{\namedref}[2]{\hyperref[#2]{#1~\ref*{#2}}}
\newcommand{\secref}{\@ifstar{\namedref{Section}}{\namedref{sec.}}}
\newcommand{\appref}{\@ifstar{\namedref{Appendix}}{\namedref{app.}}}
\newcommand{\tabref}{\@ifstar{\namedref{Table}}{\namedref{tab.}}}
\newcommand{\figref}{\@ifstar{\namedref{Figure}}{\namedref{fig.}}}
\newcommand{\Figref}{\@ifstar{\namedref{Figure}}{\namedref{Fig.}}}
\makeatother

\let\oldbib=\thebibliography
\def\thebibliography{\phantomsection\addcontentsline{toc}{section}{\refname}\oldbib}

\let\oldtoc=\tableofcontents
\def\tableofcontents{\phantomsection\addcontentsline{toc}{section}{\contentsname}\oldtoc}

\providecommand{\hypersetup}[1]{}
\providecommand{\texorpdfstring}[2]{#1}
\hypersetup{plainpages=false}
\hypersetup{pdfpagemode=UseNone}
\hypersetup{bookmarksnumbered=true}
\hypersetup{pdfstartview=FitH}
\hypersetup{colorlinks=false}
\hypersetup{citebordercolor={1 1 1}}
\hypersetup{urlbordercolor={1 1 1}}
\hypersetup{linkbordercolor={1 1 1}}

\makeatletter
\let\@keywords\@empty
\let\@subject\@empty
\providecommand{\keywords}[1]{\gdef\@keywords{#1}}
\providecommand{\subject}[1]{\gdef\@subject{#1}}
\def\thetitle{\@title}
\def\theauthor{\@author}
\def\thesubject{\@subject}
\def\thedate{\@date}
\def\thekeywords{\@keywords}
\makeatother
\AtBeginDocument{
	\hypersetup{pdftitle={\thetitle}}
	\hypersetup{pdfauthor={\theauthor}}
	\hypersetup{pdfsubject={\thesubject}}
	\hypersetup{pdfkeywords={\thekeywords}}}

\newif\ifshownote
\shownotetrue

\ifpublic\shownotefalse\fi

\ifshownote

\ifpdf\else\RequirePackage[active]{srcltx}\fi
\RequirePackage{xcolor}
\RequirePackage{ulem}

\newcommand{\remark}[2][]{{%
		\def\emph{\textsl}%
		\def\tmparga{#1}%
		\def\tmpargb{BH}\ifx\tmparga\tmpargb\normalfont\sffamily\hspace{1ex}\color[rgb]{0.5,0,0}\fi
		\def\tmpargb{FS}\ifx\tmparga\tmpargb\normalfont\sffamily\hspace{1ex}\color[rgb]{0,0.5,0}\fi
		\def\tmpargb{}\ifx\tmparga\tmpargb\color[rgb]{0,0.5,0.5}\else\normalfont\sffamily\hspace{1ex}\textbf{#1:}\fi
		#2%
		\ifx\tmparga\tmpargb\else\hspace{1ex}\fi}}

\else

\newcommand{\remark}[2][]{\ignorespaces\unskip}

\fi

\title{Two-parameter integrable deformations of the \texorpdfstring{$\AdS_3 \times \Sp^3 \times \To^4$}{AdS3 x S3 x T4} superstring}
\author{Fiona~K.~Seibold}

\begin{document}

\pdfbookmark[1]{Title Page}{title}
\thispagestyle{empty}

\vspace*{2cm}
\begin{center}
	\begingroup\Large\bfseries\thetitle\par\endgroup
	\vspace{1cm}
	
	\begingroup\theauthor\par\endgroup
	\vspace{1cm}
	
	\textit{
		Institut f\"ur Theoretische Physik,\\
		Eidgen\"ossische Technische Hochschule Z\"urich,\\
		Wolfgang-Pauli-Strasse 27, 8093 Z\"urich, Switzerland}
	\vspace{5mm}
	
	\begingroup\ttfamily\small
	fseibold@itp.phys.ethz.ch\par
	\endgroup
	\vspace{5mm}
	
	\vfill
	
	\textbf{Abstract}\vspace{5mm}
	
	
	
	\begin{minipage}{15cm}\small
For supercosets with isometry group of the form $\hat{\grp{G}} \times \hat{\grp{G}}$, the $\eta$-deformation can be generalised to a two-parameter integrable deformation with independent $q$-deformations of the two copies. We study its kappa-symmetry and write down a formula for the Ramond-Ramond fluxes. We then focus on $\hat{\grp{G}}= \grp{PSU}(1,1|2)$ and construct two supergravity backgrounds for the two-parameter integrable deformation of the $\AdS_3 \times \Sp^3 \times \To^4$ superstring, as well as explore their limits. We also construct backgrounds that are solutions of the weaker generalised supergravity equations of motion and compare them to the literature.
	\end{minipage}
	
	\vspace*{2cm}
	
\end{center}

\newpage

\tableofcontents

\section{Introduction}

The $\AdS$/CFT correspondence is a duality between string theories in $d$-dimensional Anti-de-Sitter space and Conformal Field Theories living in $d-1$ dimensions \cite{Maldacena:1997re, Witten:1998qj}. In special cases, it is possible to investigate the AdS/CFT setup thanks to techniques based on superymmetry, conformal field theory, and integrability. It is natural to ask whether the better-understood AdS/CFT instances, which are often the most (super-)symmetric ones, admit deformations that preserve their solvability. Here we consider the $\AdS_3$ superstring, which is integrable \cite{Babichenko:2009dk,Cagnazzo:2012se} and admits integrable deformations. $\AdS_3$ offers a particularly good setting, providing more control than in other dimensional cases for two main reasons. Firstly, string theories on $\AdS_3$ backgrounds are dual to two-dimensional CFTs which are often exactly solvable. Secondly, the background can be supported by a mixture of R-R and NS-NS fluxes, offering a richer landscape than in the $\AdS_5$ case. Moreover, specific points allow for particularly simple solutions which can be analysed either by integrability \cite{OhlssonSax:2011ms,Sundin:2012gc,Borsato:2012ud,Hoare:2013lja,Borsato:2014hja,Hernandez:2014eta,Baggio:2017kza} (for a review and further references see \cite{Sfondrini:2014via}) or by CFT (Wess-Zumino-Witten model) techniques \cite{Maldacena:2000hw,Maldacena:2000kv,Maldacena:2001km}. $\AdS_3$ string theories also naturally emerge when studying black-hole configurations, such as those arising in the celebrated D1-D5 system (see e.g.\ \cite{David:2002wn} for a review). Consequently, it is of interest to construct new, exactly solvable string backgrounds as deformations of the $\AdS_3$ ones.

In this paper we will focus on one particular type of integrable deformation of the type II Green-Schwarz superstring known in the literature as $\eta$-deformation or inhomogeneous Yang-Baxter deformation \cite{Delduc:2013qra,Delduc:2014kha}. This generalises the $\eta$-deformations of the principal chiral model \cite{Klimcik:2002zj, Klimcik:2008eq} and the symmetric space sigma model \cite{Delduc:2013fga}. The shape of the deformation is governed by an R-matrix solving the non-split modified classical Yang-Baxter equation on the superisometry algebra of the undeformed background. We will restrict ourselves to R-matrices of Drinfel'd Jimbo type \cite{Drinfeld:1985rx,Jimbo:1985zk,Belavin:1984}, which are defined through their action on a Cartan-Weyl basis of the superisometry algebra. More precisely, they annihilate the Cartan elements and multiply by $-i$ (respectively $+i$) the positive (respectively negative) roots. Generically, the deformed backgrounds do not solve the supergravity equations of motion \cite{Arutyunov:2013ega,Arutyunov:2015qva}. Rather, they satisfy a set of generalised supergravity equations of motion that follow from imposing kappa-symmetry or equivalently scale invariance of the model, but not its Weyl invariance \cite{Arutyunov:2015mqj, Wulff:2016tju}. Progress has been made in trying to give a good string theory interpretation to these generalised supergravity backgrounds but there are a number of remaining issues \cite{Sakamoto:2017wor,Fernandez-Melgarejo:2018wpg,Muck:2019pwj}. An elegant formula for the target-space supergeometry of the $\eta$-deformed model was derived in \cite{Borsato:2016ose}. There it was shown that in order to obtain a supergravity background the R-matrix has to satisfy the so-called unimodularity condition: the supertrace of the structure constants built out of the R-bracket should vanish. This was not the case for the R-matrix chosen in \cite{Arutyunov:2015qva}. However, superalgebras admit inequivalent Dynkin diagrams, and supergravity backgrounds for the $\eta$-deformed $\AdS_2 \times \Sp^2 \times \To^6$ and  $\AdS_5 \times \Sp^5$ superstrings were finally presented in \cite{Hoare:2018ngg}. 

This solved one of the main puzzles in the field of $\eta$-deformations, but a number remain, including the behaviour in the maximal deformation limit and its link to the undeformed \textit{mirror theory} \cite{Arutyunov:2007tc,Arutyunov:2014cra,Arutyunov:2014jfa}. Under the deformation the superisometry algebra of the model gets $q$-deformed \cite{Delduc:2014kha,Delduc:2013fga,Delduc:2016ihq}, with parameter
\[
q= e^{-\kappa/T}~, \qquad \kappa = \frac{2 \eta}{1-\eta^2}~,
\]
where $\eta$ is the strengh of the deformation and $T$ is the string tension. A conjecture for the (centrally extended) $\alg{psu}_q(2|2) \oplus \alg{psu}_q(2|2)$ invariant S-matrix, which gives a quantum deformation of the $\AdS_5 \times \Sp^5$ worldsheet S-matrix, has been given in \cite{Beisert:2008tw}. This S-matrix admits three interesting limits. First of all, sending $\kappa \rightarrow 0$ gives the S-matrix of the undeformed light-cone gauge fixed $\AdS_5 \times \Sp^5$ theory. But the $q \rightarrow 1$ limit can also be reached in another way, by first rescaling the tension $T \rightarrow \kappa^2 T$ and then sending the deformation parameter to infinity. In this maximal deformation limit one gets the S-matrix of the mirror $\AdS_5 \times \Sp^5$ theory. The $q$-deformed S-matrix thus interpolates between the S-matrices of the undeformed theory and its mirror. An interesting feature of the $q$-deformed S-matrix is that it is covariant under the mirror transformation: the mirror of the $q$-deformed S-matrix with deformation parameters $q$ is again the $q$-deformed S-matrix with deformation paramter $q'$ \cite{Arutynov:2014ota}. This behaviour is referred to as mirror duality. Another interesting case is when $q$ is pure phase, which, in a particular scaling limit, gives the S-matrix of the Pohlmeyer reduced theory \cite{Hoare:2009fs,Beisert:2010kk,Hoare:2011fj}. It is an interesting and open question whether these latter two limits are also realised at the level of the background, as well as in the light-cone gauge fixed worldsheet S-matrix. For the Pohlmeyer limit, encouraging results in this direction have been obtained \cite{Hoare:2018ngg}. However, neither the generalised supergravity background of \cite{Arutyunov:2015qva} nor the supergravity background of \cite{Hoare:2018ngg} reduce to the undeformed mirror theory in the maximal deformation limit.

To shed more light on this, as yet, unresolved problem we will consider deformations of the $\AdS_3 \times \Sp^3 \times \To^4$ superstring. The novelty in this case is that the superisometry algebra has a group-product structure $\hat{\grp{G}} \times \hat{\grp{G}}$ and thus allows for a two-parameter integrable deformation \cite{Hoare:2014oua}. The latter generalises the bi-Yang-Baxter sigma model of \cite{Klimcik:2008eq,Klimcik:2014bta} to semi-symmetric space sigma models. There are now two real deformation parameters $\eta_L$ and $\eta_R$ controlling the strength of the deformation in the left and right $\hat{\grp{G}}$ copy. The limiting case $\eta_L= \eta_R=\eta$ reduces to the usual one-parameter $\eta$-deformation. We extend the results of \cite{Borsato:2016ose} to the two-parameter deformation and write down an explicit formula for the Ramond-Ramond (R-R) fluxes in this more general setting. We then apply these results to the $\AdS_3 \times \Sp^3 \times \To^4$ superstring and present two supergravity backgrounds corresponding to two different unimodular Drinfel'd Jimbo R-matrices associated to the fully fermionic Dynkin diagram
\[
\otimes - \otimes - \otimes \qquad \otimes - \otimes -\otimes~.
\] 
The two backgrounds are similar to the ones constructed in \cite{Lunin:2014tsa}, where the metric of the two-parameter deformation of the $\AdS_3 \times \Sp^3 \times \To^4$ superstring was embedded into type II supergravity in two different ways. The two supergravity backgrounds remain distinct in the limiting $\eta_L=\eta_R$ case. Studying their limits, we will observe that they have different Pohlmeyer and maximal deformation limits. Interestingly, we find that one of the background exhibits mirror duality, while the other does not. This may give new insights into how to recover the mirror background in other cases and construct integrable supergravity backgrounds for the $\eta$-deformed $\AdS_5 \times \Sp^5$ superstring with explicit mirror duality.

The outline of this paper is as follows. In \secref{sec:twoparamdef} we review the construction of the two-parameter deformation, study its kappa-symmetry and present a closed formula for the R-R fluxes. We then extract the supergravity backgrounds for the two-parameter deformation of the $\AdS_3 \times \Sp^3 \times \To^4$ superstring in \secref{sec:sugrabackgrounds} and explore their limits. In particular we will show how mirror duality arises in this context. In \secref{sec:generalised} the results are also compared to backgrounds that are solutions to the generalised supergravity equations of motion. Finally we end with some conclusions in \secref{sec:conclusions}. Our conventions for gamma matrices and generators are given in \appref{app:conventions}.
\section{Two-parameter deformation}
\label{sec:twoparamdef}
In this section the two-parameter deformation \cite{Hoare:2014oua} of the Metsaev-Tseytlin action for supercosets with isometry group of the form $\hat{\grp{G}} \times \hat{\grp{G}}$ \cite{Babichenko:2009dk} is reviewed. The kappa-symmetry of the deformed model is studied, the supervielbein are identified and a formula for the dilaton and R-R fluxes is written down. Examples of interest in the context of $\AdS$ string backgrounds are $\hat{\grp{G}}= \grp{PSU}(1,1|2)$ for strings moving in $\AdS_3 \times \Sp^3 \times \To^4$ and $\hat{\grp{G}}= \grp{D}(2,1;\alpha)$ for strings moving in $\AdS_3 \times \Sp^3 \times \Sp^3 \times \Sp^1$, where the parameter $\alpha$ is related to the relative radii of the two three-spheres. 
\subsection{The action, equations of motion and kappa-symmetry}
\paragraph{Algebraic setting.}
We shall consider deformations of semi-symmetric space sigma models on supercosets of the type
\[
\frac{\hat{\grp{G}} \times \hat{\grp{G}}}{\grp{F}_0}~,
\]
where $\grp{F}_0$ is the bosonic diagonal subgroup of the product supergroup $\hat{\grp{F}}=\hat{\grp{G}} \times \hat{\grp{G}}$. We denote by  $\hat{\alg{g}}$ and $\hat{\alg{f}}=\hat{\alg{g}} \oplus \hat{\alg{g}}$ the Lie algebras corresponding to the supergroup $\hat{\grp{G}}$ and $\hat{\grp{F}}$ respectively. The basic Lie superalgebra $\hat{\alg{f}}$ admits a $\mathbb{Z}_4$ grading consistent with the (anti-)commutation relations,
\[
\hat{\alg{f}}=\alg{f}^{(0)} + \alg{f}^{(1)} + \alg{f}^{(2)} + \alg{f}^{(3)}~, \qquad \com{\alg{f}^{(i)}}{\alg{f^{(j)}}} \subset \alg{f}^{(i+j \mod 4)},
\]
with the grade zero subalgebra $\alg{f}^{(0)}$ being the Lie algebra of $\grp{F}_0$. The subspaces $\alg{f}^{(0)}$ and $\alg{f}^{(2)}$ have even grading, while the subspaces $\alg{f}^{(1)}$ and $\alg{f}^{(3)}$ have odd grading. For elements in the Lie algebra $\hat{\alg{f}}$ we use the standard block-diagonal matrix realisation $\mathcal X = \diag(X^L, X^R) \in \hat{\alg{f}}$ with $X^L, X^R \in \hat{\alg{g}}$, and define the supertrace $\STr[\mathcal X]= \STr[X^L]+\STr[X^R]$. Let us also introduce the projectors $P^{(i)}$ onto the subspaces $\alg{f}^{(i)}$, as well as $P_\ind{B}$ and $P_\ind{F}$, which project onto the even and odd parts of $\hat{\alg{g}}$ respectively. The $\mathbb Z_4$ grading is defined through
\[
\label{eq:grading}
P^{(0)} \mathcal X &= \mathcal X^{(0)} = \frac{1}{2} \begin{pmatrix}
P_\ind{B} (X^\ind{L}+X^\ind{R}) & 0 \\
0 & P_\ind{B}(X^\ind{L}+X^\ind{R})
\end{pmatrix}~, \\
P^{(1)} \mathcal X &= \mathcal X^{(1)} = \frac{1}{2} \begin{pmatrix}
P_\ind{F} (X^\ind{L}+i X^\ind{R}) & 0 \\
0 & -i P_\ind{F}(X^\ind{L}+i X^\ind{R})
\end{pmatrix}~, \\
P^{(2)} \mathcal X &= \mathcal X^{(2)} = \frac{1}{2} \begin{pmatrix}
P_\ind{B} (X^\ind{L}-X^\ind{R}) & 0 \\
0 & -P_\ind{B}(X^\ind{L}-X^\ind{R})
\end{pmatrix}~, \\
P^{(3)} \mathcal X &= \mathcal X^{(3)} = \frac{1}{2} \begin{pmatrix}
P_\ind{F} (X^\ind{L}-i X^\ind{R}) & 0 \\
0 & i P_\ind{F}(X^\ind{L}-i X^\ind{R})
\end{pmatrix}~.
\]
The generators $\mathcal T_\ind{A}, A=1, \dots, \dim \hat{\alg{f}}$ of the superisometry algebra $\hat{\alg{f}}$ can then be split into generators $\mathcal J_{ab}$ of grade 0, $\mathcal P_{a}$ of grade 2, as well as supercharges $\mathcal Q_{1 \alpha}$ and $\mathcal Q_{2 \alpha}$ of grades 1 and 3 respectively. Let us also define
\[
\mathcal{K}_\ind{AB} = \STr[\mathcal T_\ind{A} \mathcal T_\ind{B}] ~,
\]
as well as its inverse
\[
\mathcal{K}_\ind{AB} \widehat{\mathcal{K}}^\ind{BC} = \delta_\ind{A}^\ind{C} ~.
\]

\paragraph{Action.} The action of the two-parameter deformed semi-symmetric space sigma model for the group-valued field $g \in \hat{\grp{F}}$ depends on two real deformation parameters $\eta_L$ and $\eta_R$ and reads 
\[
\label{eq:2param_action}
\Act_{\eta_L,\eta_R}[g] = - T \int \extder^2 \sigma \, \Xi_-^{ij} \STr[g^{-1} \partial_i g \, d_- O_-^{-1} \, g^{-1} \partial_j g] ~.
\]
We assume that the fields are in the defining matrix representation. $T$ is the overall coupling constant playing the role of the effective string tension, $\extder^2 \sigma = \extder \tau \extder \sigma$ and $\Xi_\pm^{ij} = (\gamma^{ij} \pm \epsilon^{ij})/2$, where $\gamma^{ij}$ is the Weyl invariant worldsheet metric with $\gamma^{\tau \tau}<0$ and $\epsilon^{ij}$ is the Levi-Civita symbol with $\epsilon^{\tau \sigma} = 1$. The operators $d_\pm$ are defined in terms of the $\Integer_4$ projectors as 
\[
d_\pm = P^{(2)} \mp \frac{\bidef}{2} (P^{(1)} - P^{(3)})~, \qquad  \bidef = \sqrt{(1-\eta_L^2)(1-\eta_R^2)}~,
\]
and
\[
O_\pm = 1\pm \diag(\kappa_L,\kappa_R) \mathcal{R}_g d_\pm~, \qquad \kappa_L = \frac{2 \eta_L}{\bidef}~, \qquad \kappa_R = \frac{2 \eta_R}{\bidef}~.
\]
The operator $\mathcal R_g = \Ad_g^{-1} \mathcal R \Ad_g^{\vphantom{-1}}$ acts on $\mathcal X \in \hat{\alg{f}}$ as $\mathcal R_g(\mathcal X)=g^{-1} \mathcal R(g\mathcal Xg^{-1})g$.
The R-matrix $\mathcal R$ governing the shape of the deformation is antisymmetric with respect to the supertrace
\[
\STr[\mathcal R(\mathcal X) \mathcal Y]=-\STr[\mathcal X \mathcal R(\mathcal Y)] ~, \qquad \mathcal X,\mathcal Y \in \hat{\alg{f}}~,
\]
and solves the non-split modified classical Yang-Baxter equation \footnote{We use the mixed bracket notation. If the two elements in the bracket are of odd grading then the bracket is the anti-commutator. Otherwise, it is the commutator.}
\[
\mcom{\mathcal R(\mathcal X)}{\mathcal R(\mathcal Y)}-\mathcal R(\mcom{\mathcal R(\mathcal X)}{\mathcal Y} + \mcom{\mathcal X}{\mathcal R(\mathcal Y)}) = \mcom{\mathcal X}{\mathcal Y}~, \qquad \mathcal X,\mathcal Y \in \hat{\alg{f}} ~.
\]

It has been conjectured that the symmetry of this model is, at least at the classical level, given by the asymmetrical $q$-deformation \cite{Hoare:2014oua}
\[
\mathcal U_{q_L}(\hat{\grp{G}}) \times \mathcal U_{q_R}(\hat{\grp{G}}) ~, \qquad q_L = e^{- \kappa_L/T} ~, \qquad q_R = e^{- \kappa_R/T}~.
\]
It reduces to the one-parameter $\eta$-deformation if $\eta_L=\eta_R=\eta$ and to the undeformed sigma model when the deformation parameters $\eta_L=\eta_R=0$. 

To write down the equations of motion and the kappa symmetry variation it will be useful to define the deformation parameters
\[
\kappa_\pm = \frac{1}{2} (\kappa_L \pm \kappa_R) 
\]
as well as the auxiliary operator
\[
\label{eq:defRtilda}
\tilde{\mathcal R} \equiv \diag(\kappa_L, \kappa_R) \mathcal R =  (\kp \identity + \km W) \mathcal R~,
\]
where we have introduced $W=\diag(1,-1)$.
This new operator is still antisymmetric with respect to the supertrace,
\[
\STr[\tilde{\mathcal R}(\mathcal X) \mathcal Y] = - \STr[\mathcal X \tilde{\mathcal R}(\mathcal Y)]~.
\]
Furthermore, the fact that $\mathcal R$ satisfies the modified classical Yang-Baxter equation implies a similar equality for $\tilde{\mathcal R}$ but with the right hand side given by 
\[
\tilde{C} \mcom{X}{Y}~, \qquad \tilde{C}= (\kp^2 + \km^2) \identity + 2 \kp \km W~.
\]

\paragraph{Equations of motion.} Defining the one-forms $A_\pm = O_\pm^{-1} g^{-1} \extder g$ with components $A_{\pm i}$, the equations of motion corresponding to the action \eqref{eq:2param_action} can be written as $\mathcal E=0$, with
\[
\label{eq:eom}
\mathcal E = d_-\partial_i (\Xi_- A_{-})^i + d_+ \partial_i ( \Xi_+ A_{+})^i + \com{(\Xi_+ A_{+})_i}{d_- (\Xi_- A_{-})^i} + \com{(\Xi_- A_{-})_i}{d_+  (\Xi_+ A_{+})^i}~.
\]
The flatness condition for the undeformed currents $g^{-1} \extder g$ also implies $\mathcal Z =0$, with
\[
\mathcal Z &= \partial_i  (\Xi_+ A_{+})^i - \partial_i  (\Xi_- A_{-})^i + \com{ (\Xi_-A_{-})_i}{(\Xi_+ A_{+})^i} \\
&\qquad + \tilde{C} \com{d_- (\Xi_- A_{-})_i}{d_+ (\Xi_+ A_{+})^i} + \tilde{\mathcal R}_g (\mathcal E)~.
\]
\paragraph{Virasoro constraints.}
These equations of motion are supplemented by the Virasoro constraints, which are obtained by varying the action with respect to the worldsheet metric. They are equivalent to imposing a vanishing worldsheet stress tensor. To derive the Virasoro constraints we focus on the part of the action that is proportional to the Weyl-invariant metric. Its variation gives
\[
\delta_\gamma \Act_{\eta_L,\eta_R}[g] = - \frac{T}{2} \int \extder^2 \sigma \, \delta \gamma^{ij} \STr[A_{-i}^{(2)}A_{-j}^{(2)}] =  -\frac{T}{2} \int \extder^2 \sigma \, \delta \gamma^{ij} \STr[A_{+i}^{(2)}A_{+j}^{(2)}]~,
\]
from which we deduce the Virasoro constraints
\[
\STr[(\Xi_-A_{-}^{(2)})_i (\Xi_- A_{-}^{(2)})^{j}]=0~, \qquad \STr[(\Xi_+ A_{+}^{(2)})_i(\Xi_+ A_{+}^{(2)})^{j}]=0~.
\]
\paragraph{kappa-symmetry.}
In addition to reparametrisation invariance and local right-acting gauge symmetry $g \rightarrow g h, h \in \grp{F}_0$, the action \eqref{eq:2param_action} also has a local right-acting fermionic kappa symmetry reducing the number of physical fermionic degrees of freedom. 

Let us consider an infinitesimal right translation of the field $\delta_\varkappa g = g \epsilon$ with
\[
\epsilon &= (W_- - \frac{\bidef}{2} W_+ \tilde{\mathcal R}_g) \rho^{(1)} + (W_- + \frac{\bidef}{2} W_+ \tilde{\mathcal R}_g) \rho^{(3)}~, \\
 W_\pm &= 1 \pm \frac{ 2 \bidef^2 \kp \km W}{4(1+\bidef)-\bidef^2(\kp^2+\km^2)}~,
\]
where $W$ has been defined under \eqref{eq:defRtilda}, and $\rho^{(1)}$ and $\rho^{(3)}$ are yet arbitrary functions on the string worldsheet. Equivalently, this transformation can be written as
\[
\label{eq:kappa_g}
O_+^{-1}(g^{-1} \delta_\varkappa g ) = W_- (\rho^{(1)} +\rho^{(3)})~.
\]
The variation of the action with respect to the field $g$ is then
\[
\label{eq:var_action_1}
\delta_g \Act_{\eta_L,\eta_R}[g] = T \int \extder^2 \sigma \STr\Big[\rho^{(1)} P_3 (W_- + \frac{\bidef}{2} W_+ \tilde{\mathcal R}_g)(\mathcal{E}) + \rho^{(3)} P_1 (W_- - \frac{\bidef}{2} W_+ \tilde{\mathcal R}_g) (\mathcal{E})\Big]~.
\]
To simplify the expression within the supertrace we use the following combination of the equations of motion and the flatness condition
\[
P_1 (W_- - \frac{\bidef}{2} W_+ \tilde{\mathcal R}_g)(\mathcal E) + \frac{\bidef}{2} P_1 W_+ (\mathcal Z) &= -2 \bidef \com{(\Xi_+ A_{+}^{(2)})_i}{ (\Xi_- \mathcal A_{-}^{(3)})^i}~, \\
P_3 (W_- + \frac{\bidef}{2} W_+ \tilde{\mathcal R}_g)(\mathcal E) - \frac{\bidef}{2} P_1 W_+(\mathcal Z) &= -2 \bidef \com{(\Xi_-A_{-}^{(2)})_i}{ (\Xi_+\mathcal A_{+}^{(1)})^i} ~,
\]
where, for convenience, we have defined
\[
\mathcal{A}_\pm =   W_+ A_{\pm} ~.
\]
The variation of the action \eqref{eq:var_action_1} can then be rewritten
\[
\label{eq:var_action_2}
\delta_g \Act_{\eta_L,\eta_R}[g] = -2 \bidef T \int \extder^2 \sigma \STr\Big[\rho^{(1)}\com{(\Xi_-A_{-}^{(2)})_i}{ (\Xi_+\mathcal A_{+}^{(1)})^i} + \rho^{(3)} P_1 \com{(\Xi_+ A_{+}^{(2)})_i}{ (\Xi_- \mathcal A_{-}^{(3)})^i}\Big]~.
\]
As usual we then make the ansatz
\[
\rho^{(1)}= \anticom{i (\Xi_+\varkappa^{ (1)})_j}{ (\Xi_- A_{-}^{(2)})^j}~, \qquad \rho^{(3)}=\anticom{i (\Xi_- \varkappa^{(3)})_j}{ (\Xi_+ A_{+}^{(2)})^j}~,
\]
where $\varkappa^{(1)}$ and $\varkappa^{(3)}$ belong to the grade 1 and grade 3 subspaces respectively. Using the fact that the $\tau$ and $\sigma$ components of the projections are proportional to each other, for instance $(\Xi_+ A_+)^\tau \sim (\Xi_+ A_+)^\sigma$, one can show that
\[
\STr\big[\rho^{(1)} \com{(\Xi_- A_{-}^{(2)})_i}{(\Xi_+\mathcal A_{+}^{(1)})^i}\big] &= \STr \big[ (\Xi_-A_{-}^{(2)})_i (\Xi_-A_{-}^{(2)})_j \com{ (\Xi_+ \mathcal A_{+}^{ (1)})^i}{i (\Xi_+ \varkappa^{(1)})^j}\big]~, \\
\STr\big[\rho^{(3)} \com{(\Xi_+A_{+}^{(2)})_i}{(\Xi_-\mathcal A_{-}^{(3)})^i}\big] &= \STr \big[ (\Xi_+A_{+}^{(2)})_i (\Xi_+ A_{+}^{(2)})_j \com{ (\Xi_- \mathcal A_{-}^{ (3)})^i}{i (\Xi_-\varkappa^{(3)})^j}\big] ~.
\]
As discussed in \cite{Grigoriev:2007bu,Arutyunov:2009ga}, the square of an elements of grade 2 yields two terms: one is proportional to the identity while the other is proportional to the fermionic parity operator, or hypercharge, $\Upsilon_{\hat{\grp{F}}}=\diag(\Upsilon_{\hat{\grp{G}}},\Upsilon_{\hat{\grp{G}}})$, where $\Upsilon_{\hat{\grp{G}}}=\diag(1,-1)$ in the defining representation. The part proportional to the identity drops out since the supertrace of a commutator vanishes. Finally, the variation of the action coming from the variation of the field $g$ is
\[
\label{eq:kappa_field}
\delta_g \Act_{\eta_L,\eta_R}[g] = & -\frac{T \bidef }{4}  \int \extder^2 \sigma \Big(\STr\Big[(\Xi_- A_{-}^{(2)})_i (\Xi_- A_{-}^{(2)})_j\Big]\STr\Big[ \Upsilon_{\hat{\grp{F}}} \com{ (\Xi_+ \mathcal A_{+}^{ (1)})^i}{i (\Xi_+\varkappa^{(1)})^j}\Big] \\
&\qquad + \STr \Big[(\Xi_+ A_{+}^{(2)})_i (\Xi_+ A_{+}^{(2)})_j \Big] \STr \Big[\Upsilon_{\hat{\grp{F}}} \com{ (\Xi_-\mathcal A_{-}^{(3)})^i}{i (\Xi_- \varkappa^{ (3)})^j} \Big] \Big)~.
\]
This term can then be compensated by the following change in the metric,
\[
\label{eq:kappa_metric}
\delta_\varkappa \gamma^{ij} = \frac{\bidef}{2} \STr \Big[ \Upsilon_{\hat{\grp{F}}} \com{i (\Xi_+ \varkappa^{ (1)})^i}{ (\Xi_+ \mathcal A_{+}^{(1)})^j} + \Upsilon_{\hat{\grp{F}}} \com{i (\Xi_-\varkappa_{- }^{(3)})^j }{(\Xi_- \mathcal A_{-}^{(3)})^i}\Big]~,
\]
which shows that the action \eqref{eq:2param_action} is kappa-symmetric.

\paragraph{Identification of supervielbein.}
Let us now bring the kappa-symmetry transformations \eqref{eq:kappa_g} and \eqref{eq:kappa_metric} into their standard Green-Schwarz form. This in turn will allow us to identify the supervielbein of the deformed theory. 

We start by considering  the equations of motion in the fermionic sector, which, according to \eqref{eq:var_action_2}, are
\[
\label{eq:eom_fermions}
\com{(\Xi_+ A_{+}^{(2)})_i}{(\Xi_-\mathcal A_{- }^{ (3)})^i} =0~, \qquad 
\com{(\Xi_- A_{-}^{(2)})_i}{(\Xi_+\mathcal A_{+}^{(1)})^i} =0~.
\]
To compare these expressions with the equations of motion of the undeformed model we need to find a relation between $A_+^{(2)}$ and $A_-^{(2)} = P^{(2) }O_-^{-1} O_+ A_+$. Defining $M=O_-^{-1} O_+$ we find, just as for the one-parameter deformation, that $P^{(2)} M P^{(2)}$ implements a Lorentz transformation on the grade-2 subspace of the superisometry algebra,
\[
P^{(2)} M P^{(2)} = \Ad_h^{-1} P^{(2)} = P^{(2)} \Ad_h^{-1}~, \qquad h \in \grp{F}_0~.
\] 
Furthermore, 
\[M= 1- 2 P^{(2)}+2 O_-^{-1}P^{(2)}\]
also still holds. Therefore, $A_{-i}^{(2)}= \Ad_h^{-1} A_{+i}^{(2)}$ and the equations of motion \eqref{eq:eom_fermions} take the same form as the equations of motion of the undeformed model if one identifies the supervielbein as 
\[
E^{(2)} =A_+^{(2)}~, \qquad 
E^{(1)} =  \zeta \Ad_h \mathcal A_+^{(1)} ~, \qquad 
E^{(3)}  = \zeta \mathcal A_-^{(3)}~, 
\]
with normalisation \footnote{We will see later in \secref{sec:RRfluxes} that it is this normalisation that brings the torsion into its standard Green-Schwarz form.}
\[
\zeta = \frac{\sqrt{1+\bidef-\frac{\bidef^2}{4}(\kp^2+\km^2)}}{\sqrt{2}}~.
\]
In the above $E^{(2)} \equiv E^a \mathcal P_a$ is the bosonic supervielbein, and $E^{(1)} \equiv E^{1 \alpha} \mathcal Q_{1\alpha}$, $E^{(3)}\equiv E^{2 \alpha} \mathcal Q_{2 \alpha}$ are the fermionic supervielbein. 

Finally, if one identifies the supervielbein in this way, we can write the kappa transformations \eqref{eq:kappa_g} and \eqref{eq:kappa_metric}  as \footnote{The interior derivative $\intder_{\delta_\varkappa}$ is such that, for instance, $\intder_{\delta_\varkappa} O_+^{-1} (g^{-1} \extder g) = O_+^{-1} (g^{-1} \delta_\varkappa g)$.}
\[
&\intder_{\delta_\varkappa} E^{(2)}=0~, \qquad \intder_{\delta_\varkappa} E^{(1)}= \Xi_-^{ij} \anticom{i \hat{\varkappa}_i^{(1)}}{E_j^{(2)}}~, \qquad \intder_{\delta_\varkappa}E^{(3)}= \Xi_+^{ij} \anticom{i \hat{\varkappa}_i^{(3)}}{E_j^{(2)}}~,\\
 & \delta_\varkappa \gamma^{ij} = \frac{1}{2} \STr \Big[ \Upsilon_{\hat{\grp{F}}} \com{(\Xi_+ i  \hat{\varkappa}^{ (1)})^i}{(\Xi_+ E^{(1)})^j} + \Upsilon_{\hat{\grp{F}}} \com{(\Xi_- i  \hat{\varkappa}^{ (3)})^i}{(\Xi_- E^{(3)})^j}\Big]~, \\
 &\hat{\varkappa}^{(1)}= \frac{\bidef}{\zeta} \Ad_h \varkappa^{(1)}~, \qquad \hat{\varkappa}^{(3)}= \frac{\bidef}{\zeta}  \varkappa^{(3)}~.
\]
This shows that the kappa-symmetry variation takes the standard Green-Schwarz form and that the vielbein have been chosen appropriately. 

\subsection{Extracting the Ramond-Ramond fluxes}
\label{sec:RRfluxes}
In \cite{Borsato:2016ose} a formula expressing the R-R bispinor in terms of the operators $O_\pm$ appearing in the $\eta$-deformed sigma model has been written down. We would like to extend those results to the two-parameter deformation. To achieve this, we will follow the same steps as in \cite{Borsato:2016ose}, generalising when needed. The strategy is as follows. The study of the kappa-symmetry variation has allowed us to identify the supervielbein. By comparing the superspace torsion with its usual Green-Schwarz expression, the exterior derivative of the supervielbein can be linked to the spin connection $\Omega_{ab}$, NS-NS three-form $H_{abc}$, R-R bispinor $\mathcal S^{1 \alpha 2 \beta}$, as well as the dilatino and gravitino field strength superfields. In contrast to \cite{Borsato:2016ose}, we shall only be concerned with extracting the R-R fluxes and thus will only need the leading terms in the expansion in fermions. To leading order the relations take the form \footnote{We have contracted spinor indices and suppressed the $\wedge$ for readability. }
\[
\extder E^a &= -\frac{i}{2} E^1 \gamma^a E^1 - \frac{i}{2} E^2 \gamma^a E^2 - E^b \Omega_{b}{}^a~, \\
\extder E^{1\alpha} &= \frac{1}{4} (\gamma_{ab} E^2)^\alpha \Omega^{ab} - \frac{1}{8} E^a (E^1 \gamma^{bc})^\alpha H_{abc} -\frac{1}{8} E^a (E^2 \gamma_a \mathcal S^{12})^\alpha ~, \\
\extder E^{2\alpha} &= \frac{1}{4} (\gamma_{ab} E^2)^\alpha \Omega^{ab} + \frac{1}{8} E^a (E^2 \gamma^{bc})^\alpha H_{abc} -\frac{1}{8} E^a (E^1 \gamma_a \mathcal S^{12})^\alpha ~.
\]
We start by calculating the exterior derivative of $A_+$, \footnote{Some of the following results differ from \cite{Borsato:2016ose} by a sign. This comes from the fact that in our conventions the exterior derivative $\extder$ is acting from the left. We also use a different convention for the components of $n$-forms, namely $A_n = \frac{1}{n!} A_{\mu_1 \dots \mu_n} \extder X^{\mu_1} \wedge \dots \wedge \extder X^{\mu_n}$.}
\[
\extder A_+ &= - \frac{1}{2} O_+^{-1} \anticom{A_+}{A_+} - O_+^{-1} \tilde{\mathcal R}_g \anticom{A_+}{d_+ A_+} -  O_+^{-1} \tilde{\mathcal R}_g \anticom{\tilde{\mathcal R_g} d_+ A_+}{d_+ A_+}\\
&\qquad +\frac{1}{2} O_+^{-1}\anticom{\tilde{\mathcal R}_g d_+ A_+}{\tilde{\mathcal R}_g d_+ A_+} \\
&= -\frac{1}{2} O_+^{-1} \anticom{A_+}{A_+}+\frac{1}{2} O_+^{-1}  \tilde{C} \anticom{ d_+ A_+}{ d_+ A_+} - O_+^{-1} \tilde{\mathcal R}_g \anticom{A_+}{d_+ A_+}~,
\]
where in the last equation we have used the modified classical Yang Baxter equation. We also introduced the notation $\anticom{X}{Y}= X^\ind{A} \wedge Y^{\ind{B}} \mcom{\mathcal T_\ind{A}}{\mathcal T_{\ind{B}}}$ for one-forms $X=X^\ind{A} \mathcal T_\ind{A}$ and $Y=Y^\ind{B} \mathcal T_\ind{B}$.
Similarly, the result for $\extder A_-$ is
\[
\extder A_- = -\frac{1}{2} O_-^{-1} \anticom{A_-}{A_-}+\frac{1}{2}  O_-^{-1} \tilde{C}\anticom{ d_- A_-}{ d_- A_-} +  O_-^{-1} \tilde{\mathcal R}_g \anticom{A_-}{d_- A_-}~.
\]
To proceed we rewrite these expressions as
\[
\extder A_+ &= -\frac{1}{2} \anticom{A_+}{A_+} + \frac{1}{2} \tilde{C} \anticom{d_+ A_+}{d_+ A_+} - (O_+^{-1} -1)(X_+) -  O_+^{-1} \tilde{\mathcal R}_g \anticom{A_+^{(2)}}{A_+^{(2)}}~, \\
\extder A_- &= -\frac{1}{2} \anticom{A_-}{A_-} + \frac{1}{2} \tilde{C} \anticom{d_- A_-}{d_- A_-} - (O_-^{-1} -1)(X_-) +  O_-^{-1} \tilde{\mathcal R}_g \anticom{A_-^{(2)}}{ A_-^{(2)}}~,
\]
where
\[
X_\pm &= \frac{\bidef^2}{2} \kp \km W \anticom{A_\pm^{(1)}}{A_\pm^{(3)}}- \kp \km W \anticom{A_\pm^{(2)}}{A_\pm^{(2)}} \\
&\pm\frac{2}{\bidef} \left(1\pm\bidef - \frac{\bidef^2}{4} (\kp^2+\km^2)-\frac{\bidef^2}{2} \kp \km W \right) \anticom{A_\pm^{(2)}}{A_\pm^{(3)}} \\
&+\frac{1}{2} \left(1\pm\bidef-\frac{\bidef^2}{4}(\kp^2 + \km^2)\right) \anticom{A_\pm^{(1)}}{A_\pm^{(1)}} + \frac{1}{2} \left(1\mp\bidef-\frac{\bidef^2}{4}(\kp^2+\km^2)\right) \anticom{A_\pm^{(3)}}{A_\pm^{(3)}} \\
&\mp\frac{2}{\bidef} \left(1\mp\bidef - \frac{\bidef^2}{4}(\kp^2+\km^2)-\frac{\bidef^2}{2}  \kp \km W \right) \anticom{A_\pm^{(2)}}{A_\pm^{(1)}}~.
\]
Assuming that the operators $\tilde{\mathcal R}_g$ and $O_\pm$ do not mix the bosonic and fermionic sector (which is the case to leading order in the expansion), the projection onto $P^{(2)}$ gives
\[
\extder E^{(2)}&= - \frac{1}{2} \anticom{E^{(1)}}{E^{(1)}}- \frac{1}{2} \anticom{E^{(3)}}{E^{(3)}}  \\ 
&\qquad -\anticom{A_+^{(0)}}{E^{(2)}} - P_2 O_+^{-1} (\tilde{\mathcal R}_g - \kp \km W) \anticom{E^{(2)}}{E^{(2)}}~.
\]
We can thus identity the spin connection as
\[
\Omega_{ab}= -(\tilde{A})_{ab} + \frac{1}{2} E^c (2 \tilde{M}_{c[a,b]} - \tilde{M}_{ab,c})~, \quad \tilde{A} = (1-\kp \km W/2) A_+~, \quad \tilde{M}= (1 + \kp \km W)M~.
\]
Furthermore, we also find that
\[
\extder E^{(3)} &= - \anticom{A_+^{(0)}}{E^{(3)}}-2\anticom{P_0 O_-^{-1}E^{(2)}}{E^{(3)}}-  \kp \km W \anticom{\Ad_h^{-1} E^{(2)}}{E^{(3)}} \\
&\qquad + \left(1+\frac{2}{\bidef^2} - \frac{1}{2} (\kp^2 + \km^2) \right) \Ad_h^{-1} \anticom{E^{(2)}}{E^{(1)}} \\
&\qquad -\frac{2}{\bidef} \left(1+\bidef-\frac{\bidef^2}{4}(\kp^2+\km^2)\right) W_+ O_-^{-1} W_- \Ad_h^{-1} \anticom{E^{(2)}}{E^{(1)}}~,
\]
from which we obtain the NS-NS three-form and R-R bispinor
\[
\label{eq:bispinor}
H_{abc} &= 3 \tilde{\tilde{M}}_{[ab,c]}~, \qquad \tilde{\tilde{M}}= \tilde{M} +  \frac{1}{3} \kp \km W ~,\\
\mathcal S^{1 \alpha 2 \beta} &= 8 i \left[\Ad_h \left(1+\frac{1}{1-\eta_L^2} + \frac{1}{1-\eta_R^2} - 4 \tilde{O}_+^{-1}\right)\right]^{1 \alpha}_{~~~1 \gamma} \hat{K}^{1 \gamma 2 \beta}~,
\]
where
\[
\tilde{O}_\pm = (1+\tilde{\Omega}_\pm)^{-1}~, \qquad \tilde{\Omega}_\pm = \pm \diag\left(\frac{\eta_L}{1-\eta_L^2}, \frac{\eta_R}{1-\eta_R^2}\right) \mathcal R_g d_\pm~.
\]
In the above expression $g$ is a purely bosonic group-valued field. We immediately see that when $\eta_L = \eta_R=\eta$ this formula reduces to the one of \cite{Borsato:2016ose}. 

The bispinor \eqref{eq:bispinor} should then be compared to the familiar expression
\[
\label{eq:bispinorTH}
\mathcal{S} = -i \sigma_2 \gamma^a \RRF_{a} - \frac{1}{3 !} \sigma_1 \gamma^{abc} \RRF_{abc} - \frac{1}{2 \cdot 5!} i \sigma_2 \gamma^{abcde} \RRF_{abcde} ~,
\]
valid for a type IIB supergravity background and written in terms of $16 \times 16$ chiral gamma matrices. Comparing the two expressions gives the one-form $\RRF_1$, three-form $\RRF_3$ and five-form $\RRF_5$. Moreover, for standard supergravity backgrounds the R-R fluxes are 
\[
F_n = e^{-\Phi} \mathcal F_n~,
\]
where $\Phi$ is the dilaton. In analogy with the one-parameter deformation we postulate that the latter is given by
\[
\label{eq:dilaton}
e^{-2 \Phi} = e^{-2 \Phi_0} \sdet(O_+)~,
\]
which will be supported by specific examples. We also define the R-R potentials $C_n$, defined though
\[
F_n = \extder C_{n-1} + H \wedge C_{n-3}~.
\]
\paragraph{Condition for Weyl invariance.}
In \cite{Borsato:2016ose}, a condition on the R-matrix for the one-parameter $\eta$-deformation to be a standard supergravity solution was given. This is the unimodularity property
\[\label{eq:unimodularity}
\widehat{\mathcal{K}}^\ind{AB} \STr[\mcom{\mathcal T_\ind{A}}{R(\mathcal T_\ind{B})} \mathcal Z]=0 ~, \qquad \forall \mathcal Z \in \alg{f} ~.
\]
For Lie superalgebras of the type that we are considering in this paper, this unimodularity condition is equivalent to the vanishing of the supertrace of the structure constants associated to the R-bracket \cite{Hoare:2018ngg}
\[\label{eq:superrbracket}
\mcom{\mathcal X}{\mathcal Y}_{\mathcal R} = \mcom{\mathcal X}{\mathcal R(\mathcal Y)} + \mcom{\mathcal R(\mathcal X)}{\mathcal Y} ~, \qquad \mathcal X,\mathcal Y \in \alg{f} ~,
\]
which is the generalisation to superalgebras of the R-bracket \cite{SemenovTianShansky:1983ik}. For the two-parameter deformation it is natural to postulate that if the deformation is governed by a unimodular R-matrix then the theory will be Weyl invariant and the background will solve the standard supergravity equations of motion. While our results give arguments in favour of this claim we will not provide a proof of it. \footnote{Since the one-parameter deformation is a particular case of the two-parameter deformation, the unimodularity condition \eqref{eq:unimodularity} is a necessary condition (although see \cite{Wulff:2018aku,Borsato:2018spz}) to have a supergravity solution. In order to prove that it is also a sufficient condition one would need to calculate the dilatinos $\chi_{I \alpha}$ and check that they match the spinor derivatives of the dilaton, $\nabla_{I \alpha} \Phi$. This calculation requires going to higher order in fermions and hence we do not perform it here.}
\remark{I do not have a proof that this unimodularity condition also extends to the two-parameter case so should be careful with phrasing. To get a necessary condition for having supergravity then one needs the spinor derivative $\nabla_{I \alpha} \Phi$ and impose that it is zero. In principle easy calculation but need to know the exterior derivative of log of superdet (should be related to supertrace but...) }
\section{Supergravity backgrounds for the two-parameter deformation of the \texorpdfstring{$\AdS_3 \times \Sp^3 \times \To^4$}{AdS3 x S3 x T4} superstring}
\label{sec:sugrabackgrounds}
In this section we shall focus on the case $\hat{\grp{G}}=\grp{PSU}(1,1|2)$ and consider deformations of the semi-symmetric space sigma model based on the supercoset
\[
\label{eq:supercosetAdS3}
\frac{\grp{PSU}(1,1|2) \times \grp{PSU}(1,1|2) }{\grp{SU}(1,1) \times \grp{SU(2)}}~.
\]
We choose unimodular R-matrices and construct the embedding of the 6-dimensional backgrounds in 10 dimensions with the remaining compact dimensions given by a four-torus. This then gives supergravity backgrounds for the two-parameter deformation of the $\AdS_3 \times \Sp^3 \times \To^4$ superstring. 
\subsection{Choice of R-matrix}
The superisometry algebra of the $\AdS_3 \times \Sp^3$ semi-symmetric space is $\alg{psu}(1,1|2)_L \oplus \alg{psu}(1,1|2)_R$, with two copies of the $\alg{psu}(1,1|2)$ superalgebra that we shall refer to as the left copy (subscript $L$) and the right copy (subscript $R$). We consider deformations governed by R-matrices of the type $\mathcal R = \diag(R_L, R_R)$, where $R_L$ and $R_R$ are Drinfel'd Jimbo R-matrices satisfying the non-split modified classical Yang-Baxter equation on $\alg{psu}(1,1|2)_L$ and $\alg{psu}(1,1|2)_R$ respectively. 

As already discussed in \cite{Hoare:2018ngg}, the complexified algebra $\alg{(p)sl}(2|2)$ admits three inequivalent Dynkin diagrams, $\Circle - \otimes -\Circle$, $\otimes - \Circle - \otimes$ and $\otimes - \otimes - \otimes$, each of which can be realised by a different choice of simple roots. The associated R-matrices generically lead to inequivalent backgrounds. In particular, while R-matrices associated to the fully fermionic Dynkin diagram $\otimes - \otimes - \otimes$ are unimodular and hence are expected to give rise to supergravity backgrounds, this is not the case for R-matrices associated with the other two Dynkin diagrams. 

Let us recall how to construct the different R-matrices associated with the superalgebra $\alg{psu}(1,1|2)$. We consider a realisation of the $\alg{psu}(1,1|2)$ algebra in terms of $4 \times 4$ supermatrices. The $2 \times 2$ upper left block generates the $\alg{su}(1,1)$ subalgebra, while the lower right block generates the $\alg{su}(2)$ subalgebra. The remaining entries are fermionic; see \appref{app:conventions} for the conventions we use. The various R-matrices can then be constructed by considering permutations of 4 elements.  Namely, starting from a reference R-matrix $R_0$ associated with the distinguished Dynkin diagram $\Circle - \otimes -\Circle$ and whose explicit action on an element $M \in \alg{psu}(1,1|2)$ is given by
\[
\label{eq:R0}
R_0(M)_{ij} = -i  \epsilon_{ij} M_{ij}~, \qquad \epsilon = \begin{pmatrix}
0 & +1 & + 1 & +1 \\
-1 & 0 & +1 & + 1 \\
- 1 & -1 & 0 & +1 \\
-1 & - 1 & -1 & 0
\end{pmatrix}~,
\]
we act with the permutation matrix $P_{ij} = \delta_{\mathds{P}(i)j}$ to obtain the new R-matrix
\[\label{eq:permutedrmatrix}
R_{\mathds{P}} = \Ad_P^{-1} R_0 \Ad_P^{\vphantom{-1}} ~.
\]
Of the $24$ possible permutations, 8 give rise to unimodular R-matrices. If one further demands the action on $\alg{su}(1,1)$ and $\alg{su}(2)$ is left invariant (that is to say, the new R-matrix  $R_{\mathds{P}}$ and the reference R-matrix $R_0$ have the same action on the bosonic generators) then only the permutations that do not exchange the order of $\{1,2\}$ and $\{3,4\}$ need to be considered. This leads to the two following unimodular R-matrices
\[
\mathds{P}_1 = \begin{pmatrix}
1 & 2 & 3 & 4 \\
1 & 3 & 2 & 4
\end{pmatrix}~, \qquad R_{\mathds{P}_1}(M)_{ij} = -i  \epsilon_{ij} M_{ij}~, \qquad \epsilon = \begin{pmatrix}
0 & +1 & + 1 & +1 \\
-1 & 0 & -1 & + 1 \\
- 1 & +1 & 0 & +1 \\
-1 & - 1 & -1 & 0
\end{pmatrix}~, \\
\mathds{P}_2 = \begin{pmatrix}
1 & 2 & 3 & 4 \\
3 & 1 & 4 & 2
\end{pmatrix}~, \qquad R_{\mathds{P}_2}(M)_{ij} = -i  \epsilon_{ij} M_{ij}~, \qquad \epsilon = \begin{pmatrix}
0 & +1 & - 1 & +1 \\
-1 & 0 & -1 & - 1 \\
+ 1 & +1 & 0 & +1 \\
-1 & + 1 & -1 & 0
\end{pmatrix}~.
\]
In the $\AdS_2 \times \Sp^2$ case examined in \cite{Hoare:2018ngg} these two R-matrices were not considered inequivalent, as the backgrounds are related to each other by analytical continuation. 

For the $\AdS_3 \times \Sp^3$ case there are two copies of the $\alg{psu}(1,1|2)$ superalgebra. We focus our attention on unimodular R-matrices $\mathcal R=\diag(R_L, R_R)$ associated to the completely fermionic Dynkin diagram,
\[
(\otimes - \otimes - \otimes )_L \qquad (\otimes - \otimes - \otimes )_R~.
\]
The two R-matrices $R_L$ and $R_R$ do not need to be identical and one can take different sets of positive and negative roots in the two copies of $\alg{psu}(1,1|2)$. We restrict our attention to R-matrices with the desired action on the bosonic generators and thus $R_L$ and $R_R$ can be $R_{\mathds{P}_1}$ or $R_{\mathds{P}_2}$. This then leads to four different unimodular R-matrices on the $\alg{psu}(1,1|2)_L \oplus \alg{psu}(1,1|2)_R$ superisometry algebra of the $\AdS_3 \times \Sp^3$ semi-symmetric space, \footnote{Here by convention we use opposite signs in the left and right sectors. Of course, for the two-parameter deformation the combination that enters the action is $\diag(\eta_L R_L, \eta_R R_R)$ and since $\eta_L$ and $\eta_R$ are arbitrary one can always reabsorb the minus sign by sending $\eta_R \rightarrow - \eta_R$. Our convention makes it easier to compare our results with \cite{Hoare:2014oua}, where the bosonic background has been obtained for this choice of R-matrix.}
\[
\mathcal{R}_1&=\diag(R_{\mathds{P}_1},-R_{\mathds{P}_1})~, &\qquad \mathcal{R}_2&=\diag(R_{\mathds{P}_1},-R_{\mathds{P}_2})~,  \\ \mathcal{R}_3&=\diag(R_{\mathds{P}_2},-R_{\mathds{P}_1})~, &\qquad \mathcal{R}_4&=\diag(R_{\mathds{P}_2},-R_{\mathds{P}_2})~.
\]
It transpires that the R-matrices $\mathcal{R}_1$ and $\mathcal{R}_4$, as well as $\mathcal{R}_2$ and $\mathcal{R}_3$, give rise to equivalent backgrounds related by analytic continuations and in the following we shall only consider the two inequivalent unimodular R-matrices
\[
\mathcal{R}_1=\diag(R_{\mathds{P}_1},-R_{\mathds{P}_1})~, \qquad \mathcal{R}_2= \diag(R_{\mathds{P}_1},-R_{\mathds{P}_2})~.
\]
By virtue of their unimodularity we expect the corresponding deformed backgrounds to solve the supergravity equations of motion for arbitrary deformation parameters $\eta_L$ and $\eta_R$.

\subsection{Supergravity backgrounds} 
\label{sec:supergravity_backgrounds}
In this section we extract the supergravity backgrounds corresponding to the unimodular R-matrices $\mathcal{R}_1$ and $\mathcal{R}_2$ by using the formula \eqref{eq:bispinor} for the R-R fluxes and \eqref{eq:dilaton} for the dilaton. We choose the following parametrisation for the gauge-fixed group-valued field  $g \in \hat{\grp{F}}$ ,
\begin{align}
\label{eq:param}
g &= \diag(g_L,g_R)~, \nln
g_L &= \begin{pmatrix}
\exp \left[ \frac{i}{2} (t+\psi) \sigma_3 \right] \exp \left[ \frac{1}{2}  \arsinh(\rho) \sigma_1 \right] & 0 \\
0 & \exp \left[\frac{i}{2} (\varphi+\phi) \sigma_3 \right] \exp\left[\frac{i}{2}  \arcsin(r) \sigma_1\right]
\end{pmatrix}~, \\
g_R &= \begin{pmatrix}
\exp\left[-\frac{i}{2} (t-\psi) \sigma_3 \right] \exp\left[-\frac{1}{2} \arsinh(\rho) \sigma_1\right]& 0 \\
0 & \exp\left[-\frac{i}{2} (\varphi-\phi) \sigma_3\right] \exp\left[-\frac{i}{2}  \arcsin(r) \sigma_1\right] \nn
\end{pmatrix}~.
\end{align}

The bosonic background is common to the two choices of R-matrix, with the metric and closed B-field given by
\begin{align}
\label{eq:backbi1bos}
\extder s^2 &= \frac{1}{F(\rho)} \left[-[1+\rho^2][1+\kappa_-^2 (1+\rho^2)] \extder t^2 + \frac{\extder \rho^2}{1+\rho^2} + \rho^2 [1-\kappa_+^2 \rho^2]\extder \psi^2 + 2 \kappa_- \kappa_+ \rho^2[1+\rho^2] \extder t \extder \psi \right] \nln
&+ \frac{1}{\tilde{F}(r)} \left[\phantom{-}[1-r^2][1+\kappa_-^2 (1-r^2)] \extder \varphi^2 + \frac{\extder r^2}{1-r^2} + r^2 [1+\kappa_+^2 r^2]\extder \phi^2 + 2 \kappa_- \kappa_+ r^2[1-r^2] \extder \varphi \extder \phi \right] \nln
&+ \extder x^i \extder x^i~,\\
B &=\frac{\rho}{F(\rho)} (\kappa_+ \extder t \wedge \extder \rho + \kappa_- \extder \rho \wedge \extder \psi)+ \frac{ r}{\tilde{F}(r)} ( \kappa_+  \extder \varphi \wedge \extder r + \kappa_- \extder r \wedge \extder \phi) \nn~,
\end{align}
where
\[
F(\rho)= 1+\kappa_-^2(1+\rho^2)-\kappa_+^2\rho^2~, \qquad \tilde{F}(r)= 1+\kappa_-^2(1-r^2)+\kappa_+^2r^2~.
\]
The dilaton and R-R sector depend on the choice of R-matrix. The background corresponding to the R-matrix $\mathcal{R}_1=\diag(R_{\mathds{P}_1},-R_{\mathds{P}_1})$ is
\begin{align}
\label{eq:backbi1}
e^{-2 \Phi} &= e^{-2 \Phi_0} \frac{F(\rho)\tilde{F}(r)}{P(\rho,r)^2}~, \qquad P(\rho,r) = 1-\kappa_+^2 (\rho^2-r^2-\rho^2 r^2)+\kappa_-^2(1+\rho^2)(1-r^2)~, \\
C_2 &= - \sqrt{\frac{1+\kappa_+^2}{1+\kappa_-^2}} \frac{e^{-\Phi_0}}{P(\rho,r)} \big[ \rho^2 \extder t \wedge \extder \psi + r^2 \extder \varphi \wedge \extder \phi  + \kappa_-^2(1+\rho^2) r^2 \extder t \wedge \extder \phi \nln
&\qquad- \kappa_-^2 \rho^2 (1-r^2) \extder \psi \wedge \extder \varphi + \kappa_+ \kappa_- (\rho^2-r^2-\rho^2 r^2) \extder t \wedge \extder \varphi - \kappa_+ \kappa_- \rho^2 r^2 \extder \psi \wedge \extder \phi \big]~,\nln
C_4 &= - \sqrt{\frac{1+\kappa_+^2}{1+\kappa_-^2}} \frac{e^{-\Phi_0}}{P(\rho,r)} \big[ \kappa_- \rho^2 \extder t \wedge \extder \psi + \kappa_- r^2 \extder \varphi \wedge \extder \phi - \kappa_- (1+\rho^2) r^2 \extder t \wedge \extder \phi \nln
&\qquad+ \kappa_- \rho^2 (1-r^2) \extder \psi \wedge \extder \varphi - \kappa_+ (\rho^2-r^2-\rho^2 r^2) \extder t \wedge \extder \varphi + \kappa_+ \rho^2 r^2 \extder \psi \wedge \extder \phi \Big] \wedge J_2~,\nn
\end{align}
while the background corresponding to the R-matrix   $\mathcal{R}_2=\diag(R_{\mathds{P}_1},-R_{\mathds{P}_2})$ is
\begin{align}
\label{eq:backbi2}
e^{-2 \Phi} &= e^{-2 \Phi_0} \frac{F(\rho)\tilde{F}(r)}{P(\rho,r)^2}~, \qquad P(\rho,r) = 1-\kappa_+^2 \rho^2 r^2+\kappa_-^2(1+\rho^2 r^2)~, \\
C_2 &= - \sqrt{\frac{1+\kappa_-^2}{1+\kappa_+^2}} \frac{e^{-\Phi_0}}{P(\rho,r)} \big[ (1+\rho^2) \extder t \wedge \extder \psi -(1-r^2) \extder \varphi \wedge \extder \phi  + \kappa_+^2 (1+\rho^2) r^2 \extder t \wedge \extder \phi \nln
&\qquad- \kappa_+^2 \rho^2(1-r^2) \extder \psi \wedge \extder \varphi + \kappa_+ \kappa_- (1+\rho^2) (1-r^2) \extder t \wedge \extder \varphi - \kappa_+ \kappa_- (1+\rho^2 r^2) \extder \psi \wedge \extder \phi  \big]~,\nln
C_4 &= - \sqrt{\frac{1+\kappa_-^2}{1+\kappa_+^2}} \frac{e^{-\Phi_0}}{P(\rho,r)} \big[ \kappa_+  (1+\rho^2) \extder t \wedge \extder \psi - \kappa_+ (1-r^2) \extder \varphi \wedge \extder \phi - \kappa_+ (1+\rho^2) r^2 \extder t \wedge \extder \phi \nln
&\qquad+ \kappa_+ \rho^2 (1-r^2) \extder \psi \wedge \extder \varphi - \kappa_-(1+\rho^2) (1-r^2) \extder t \wedge \extder \varphi  + \kappa_- (1+\rho^2 r^2) \extder \psi \wedge \extder \phi  \Big] \wedge J_2~. \nn
\end{align}
The Kähler form on the torus is $J_2= \extder x^1 \wedge \extder x^2 - \extder x^3 \wedge \extder x^4$. Due to their gauge symmetries the expressions for the potentials $C_2$ and $C_4$ are not unique. The form chosen here makes it manifest that in the $\kappa_\pm \rightarrow 0$ limit one recovers the undeformed background, with constant dilaton and a three-form $F_3 = \extder C_2$ proportional to the volume form on $\AdS_3 \times \Sp^3$. Let us also notice that $C_2$ and $C_4$ only contain even and odd powers of the deformation parameters $\kappa_\pm$ respectively. Since the bosonic roots are not simple the R-R fluxes mix the $\AdS_3$ and $\Sp^3$ coordinates in a non-trivial way. As expected, these two backgrounds satisfy the standard supergravity equations of motion.

\begin{figure} \centering
	\includegraphics[scale=1]{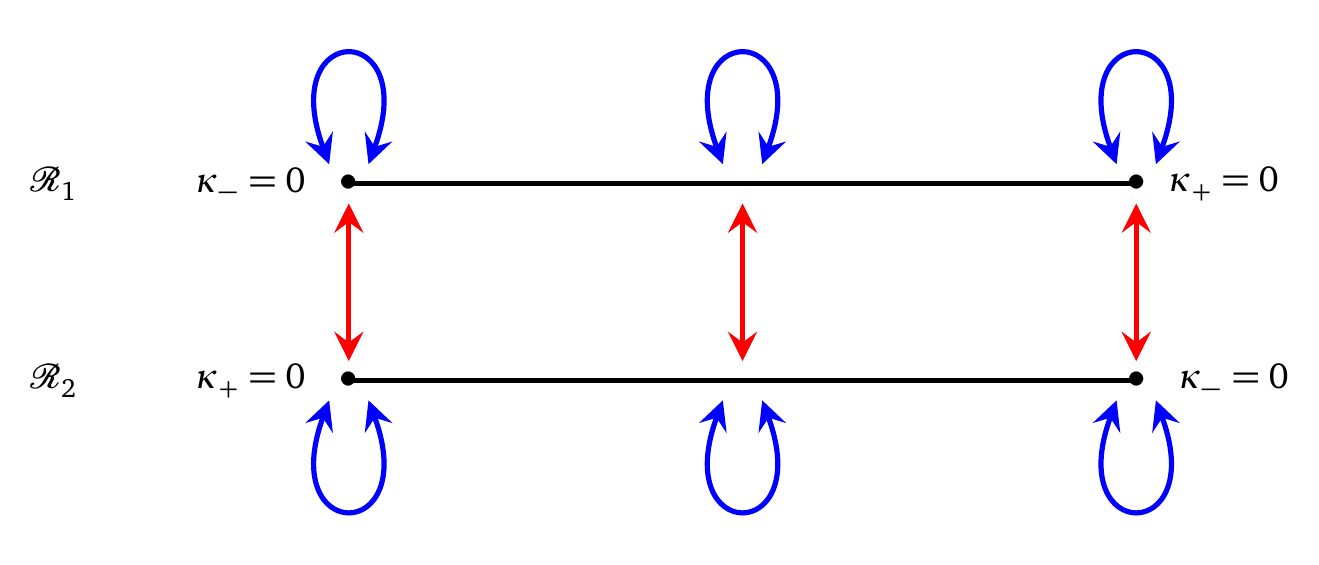}
	\caption{The two supergravity backgrounds. The top and bottom lines represent the solutions corresponding to the R-matrices $\mathcal R_1$ and $\mathcal R_2$ respectively. At the four end-points, one of the deformation parameters, $\kappa_+$ or $\kappa_-$, is set to zero. The backgrounds are left invariant under the redefinition \protect\eqref{eq:transfo1} (blue arrows), while the transformation \protect\eqref{eq:transfo2} swaps the two solutions (red arrows). }
	\label{fig:transformations}
\end{figure}

\paragraph{Relation between the two supergravity backgrounds.}
It has been observed \cite{Lunin:2014tsa, Hoare:2014oua} that the metric and B-field of the two-parameter deformed model are left invariant under the formal transformations
\[
\label{eq:transfo1}
\rho \rightarrow \frac{i \sqrt{1+\kappa_-^2}\sqrt{1+\rho^2}}{\sqrt{F(\rho)}}~,  \qquad 
r\rightarrow \frac{\sqrt{1+\kappa_-^2} \sqrt{1-r^2}}{\sqrt{F(r)}}~, \qquad t\leftrightarrow \psi~, \qquad \varphi\leftrightarrow \phi~, 
\]
and
\[
\label{eq:transfo2}
\rho \rightarrow i \sqrt{1+\rho^2}~,  \qquad 
r\rightarrow  \sqrt{1-r^2}~, \qquad t\leftrightarrow \psi~, \qquad \varphi\leftrightarrow \phi~, \qquad \kappa_+ \leftrightarrow \kappa_-~.
\]
While these two transformations involve an analytic continuation in $\rho$, it is possible to combine them to find a real transformation
\[
\label{eq:transfo3}
\kappa_+ \leftrightarrow \kappa_-~, \qquad \rho \rightarrow \frac{\sqrt{1+\kappa_-^2} \rho}{\sqrt{1+\kappa_-^2(1+\rho^2)-\kappa_+^2 \rho^2}}~,  \qquad 
r\rightarrow \frac{\sqrt{1+\kappa_-^2} r}{\sqrt{1+\kappa_-^2(1-r^2)+\kappa_+^2 r^2}}~, 
\] 
where we first interchange $\kappa_+$ and $\kappa_-$ and then do the redefinition of $\rho$ and $r$. What happens to the R-R sector under these transformations is summarised in \figref{fig:transformations}. The backgrounds \eqref{eq:backbi1} and \eqref{eq:backbi2}, including the dilaton and the R-R fluxes, are invariant under the map \eqref{eq:transfo1}. However, invariance is broken by the second set of transformations \eqref{eq:transfo2}, whose effect is to exchange \eqref{eq:backbi1} and \eqref{eq:backbi2}. Thus, also the real transformations \eqref{eq:transfo3} do not leave the supergravity backgrounds invariant but instead map between them.

It is only in the special case $\kappa_+^2=\kappa_-^2$, reached when one of the deformation parameters, either in the left or right sector, is set to zero, i.e.\  $\eta_L=0$ or $\eta_R=0$, that the background remains invariant under all three maps. In that case the two backgrounds \eqref{eq:backbi1} and \eqref{eq:backbi2} are equal, with constant dilaton.

\paragraph{Relation to previously studied supergravity backgrounds.}
In \cite{Lunin:2014tsa}, two solutions to the supergravity equations of motion with bosonic background \eqref{eq:backbi1bos} and supported by a R-R three-form flux have been constructed. Using the same terminology as in the aforementioned paper we shall refer to these two backgrounds as the $a=0$ and $a=1$ solutions. \footnote{In \cite{Lunin:2014tsa} the authors started by solving the supergravity equations of motion perturbatively in $\kappa_+$ and $\kappa_-$ and then generalized to arbitrary deformation parameter. The constant $a$ parametrizes the different solutions.} We then observe that \eqref{eq:backbi1} and  \eqref{eq:backbi2} are the analogues of the $a=1$ and $a=0$ solution of \cite{Lunin:2014tsa} respectively. The dilatons are indeed exactly the same, but we find slightly different R-R fluxes. This is a consequence of the fact that the authors of \cite{Lunin:2014tsa} were seeking solutions of the supergravity equations of motion that are only supported by a three-form flux, while the deformed backgrounds possess both a three-form and a five-form flux (or two different three-forms upon dimensional reduction). Accordingly, the three-form of \cite{Lunin:2014tsa} mixes even and odd powers of the deformation parameters $\kappa_\pm$. It is not clear if one can recover the backgrounds of \cite{Lunin:2014tsa} by performing a sequence of dualities (e.g.\ T-dualities on the torus $\To^4$), and thus their integrability remains an open question.

\subsection{Limits of the supergravity backgrounds}
Let us now explore some limits of the supergravity backgrounds of \secref{sec:supergravity_backgrounds}. We start by considering the plane-wave limit corresponding to zooming into the geometry seen by a particle moving on a light-like geodesic along a great circle of the (deformed) three-sphere. This limit can be taken for both backgrounds and for arbitrary deformation parameters $\kappa_\pm$. We also study the Pohlmeyer ($\kappa \rightarrow i$) and maximal deformation limit ($\kappa \rightarrow \infty$) of the two supergravity backgrounds when $\kappa_+=\kappa$ and $\kappa_-=0$, which corresponds to the one-parameter $\eta$-deformation of the $\AdS_3 \times \Sp^3 \times \To^4$ superstring.
\paragraph{Plane-wave limit.}
We consider the trajectory parametrised by $\tau$ along
\[
t = t(\tau)~, \qquad \rho=0~, \qquad \psi=\psi(\tau)~, \qquad \varphi=\varphi(\tau)~, \qquad r=0~, \qquad \phi=\phi(\tau)~.
\] 
Using the metric of the deformed theory, we find that a relativistic particle moving along this geodesic has action
\[
\Act = \frac{1}{2} \int \extder \tau e^{-1} \left[-\left(\frac{\extder t}{\extder \tau}\right)^2+\left(\frac{\extder \varphi}{\extder \tau}\right)^2\right]~,
\]
where $e$ is the einbein. Clearly $t=\mu \tau$, $\varphi=\mu \tau$ is a solution of the equations of motion, where we have introduced a mass scale $\mu$ in order to preserve the dimensionality of the coordinates. In order to study the geometry near this trajectory we make the transformations \cite{Berenstein:2002jq,Blau:2002dy} \footnote{The additional factor $\sqrt{1+\kappa_-^2}$ in the rescaling of $\rho$ and $r$ brings the pp-wave metric into its canonical form.}
\[
\label{eq:transfo_ppwave}
t = \mu x^+ + \frac{x^-}{\mu L^2}~, \qquad \varphi = \mu x^+ - \frac{x^-}{\mu L^2}~, \qquad \rho \rightarrow \frac{\rho \sqrt{1+\kappa_-^2}}{L}~, \qquad r \rightarrow \frac{r \sqrt{1+\kappa_-^2}}{L}~.
\]
as well as $x^i  \rightarrow x^i/L$, $T \rightarrow L^2 T$. 

In the limit $L \rightarrow \infty$ both backgrounds of \secref{sec:supergravity_backgrounds} have the same plane-wave form  \footnote{One should keep in mind that the string tension is rescaled $T \rightarrow L^2 T$ under these transformations, and that the potentials scale with the tension as $C_n \sim  T^{n/2}$, $n=0,2,4$.} 
\begin{align}
\label{eq:ppwave}
\extder s^2 &= -4 \extder x^+ \extder x^- - \mu^2 (1+\kappa_+^2)(1 + \kappa_-^2) (\rho^2 + r^2) (\extder x^+)^2 \nln
&\qquad + \extder \rho^2 + \extder r^2 + \rho^2 (\extder \psi+\kappa_+ \kappa_- \mu \extder x^+)^2 + r^2 (\extder \phi+ \kappa_+ \kappa_- \mu \extder x^+)^2 + \extder x^i \extder x^i~,  \\
B&= \rho ( \mu \kappa_+ \extder x^+ \wedge \extder \rho + \kappa_- \extder \rho \wedge \extder \psi)+ r(\mu \kappa_+ \extder x^+ \wedge \extder r + \kappa_- \extder r \wedge \phi)~, \nln
e^{-2\Phi} &= e^{-2\Phi_0} ~, \qquad C_2 = -e^{-\Phi_0} \sqrt{(1+\kappa_+^2)(1+\kappa_-^2)} \mu (\rho^2   \extder x^+ \wedge \extder \psi + r^2  \extder x^+ \wedge \extder \phi)~, \qquad C_4 = 0 \nn~.
\end{align}
Discarding the total derivative B-field we observe that in the plane-wave limit, similarly to the one-parameter deformation of the $\AdS_2 \times \Sp^2 \times \To^4$ and $\AdS_5 \times \Sp^5$ superstrings \cite{Roychowdhury:2018qsz,Hoare:2018ngg}, the deformation only enters the plane-wave background through a rescaling of the mass parameter $\mu$. 

\paragraph{Pohlmeyer limit.}
The limit $\kappa_+ = \kappa \rightarrow i, \kappa_-=0$ is interesting due to its relation to the Pohlmeyer reduced model of the undeformed $\AdS_3 \times \Sp^3 \times \To^4$ superstring \cite{Hoare:2014pna}. To take into account the fact that the string tension goes to zero in that limit we rescale, or more precisely twist, the coordinates $t$ and $\varphi$,
\[
\label{eq:transfo_Pohlmeyer}
t = \frac{\mu x^+}{\epsilon} + \frac{\epsilon x^-}{\mu}~, \qquad \varphi = \frac{\mu x^+}{\epsilon} - \frac{\epsilon x^- }{\mu}~, \qquad \kappa = i \sqrt{1-\epsilon^2}~,
\]
and take $\epsilon \rightarrow 0^+$. Only \eqref{eq:backbi1} is finite and real in this limit, giving the pp-wave background 
\[
\label{eq:Pohlmeyer}
\extder s^2 &= -4 \extder x^+ \extder x^- - \mu^2\left(\frac{\rho^2}{1+\rho^2} + \frac{r^2}{1-r^2} \right)(\extder x^+)^2 \\&\qquad+ \frac{\extder \rho^2}{(1+\rho^2)^2} + \frac{\extder r^2}{(1-r^2)^2} + \rho^2 \extder \psi^2 + r^2 \extder \phi^2 + \extder x^i \extder x^i~, \\
e^{-2 \Phi}&= e^{-2 \Phi_0} \frac{1}{(1+\rho^2)(1-r^2)}~, \\
 C_2 &= -\frac{\mu e^{-\Phi_0}}{(1+\rho^2)(1-r^2)}(\rho^2 \extder x^+ \wedge \extder \psi + r^2 \extder x^+ \wedge \extder \phi)~, \qquad C_4 =0~.
\]
We have not included the B-field, which is a divergent closed two-form with no finite contribution. This pp-wave background matches the one constructed in \cite{Hoare:2014pna}. Its light-cone gauge fixing gives the Pohlmeyer reduced theory for strings moving in undeformed $\AdS_3 \times \Sp^3$, which was constructed in \cite{Grigoriev:2008jq}. The bosonic part of the reduced theory is given by the sum of the complex sine-Gordon model and its sinh-Gordon counterpart. 
Taking the same limit but without twisting the coordinates $t$ and $\varphi$ gives the same expression but with mass $\mu=0$.

\paragraph{Maximal deformation limit.}
Another interesting limit is when the deformation parameter goes to infinity. More precisely, the maximal deformation limit \cite{Arutyunov:2014cra,Pachol:2015mfa} is given by first rescaling 
\[
 t \rightarrow \frac{t}{\kappa}~, \qquad \rho \rightarrow \frac{\rho}{\kappa}~, \qquad \phi \rightarrow \frac{\phi}{\kappa}~, \qquad r \rightarrow \frac{r}{\kappa}~,  \qquad x^i \rightarrow \frac{x^i}{\kappa} \qquad T \rightarrow \kappa^2 T~,
\]
and then taking the limit $\kappa \rightarrow \infty$. In this limit the metric and B-field become
\[
\label{eq:metric_maxdef}
\extder s^2 &= \frac{1}{1-\rho^2} (-\extder t ^2 + \extder \rho^2 ) + \rho^2 \extder \psi^2 + \frac{1}{1+r^2} (\extder \varphi^2 + \extder r^2) + r^2 \extder \phi^2 + \extder x^i \extder x^i~, \\
B &= \frac{\rho}{1-\rho^2} \extder t \wedge \extder \rho + \frac{r}{1+r^2} \extder \varphi \wedge \extder r~.
\]
The two supergravity backgrounds are both finite but remain different in this limit. The maximal deformation limit of \eqref{eq:backbi1} is
\[
\label{eq:backbi1_maxdef}
e^{-2 \Phi} &= e^{-2 \Phi_0} \frac{(1-\rho^2) (1+r^2)}{(1-\rho^2+r^2)^2}~, \qquad
C_2 = -\frac{e^{-\Phi_0}}{1-\rho^2+r^2} (\rho^2 \extder t \wedge \extder \psi + r^2 \extder \varphi \wedge \extder \phi )~, \\
C_4 &= \frac{e^{-\Phi_0}}{1-\rho^2+r^2} ((\rho^2-r^2) \extder t \wedge \extder \varphi - \rho^2 r^2 \extder \psi \wedge \extder \phi ) ~.
\]
while the maximal deformation limit of \eqref{eq:backbi2} is
\[
\label{eq:backbi2_maxdef}
e^{-2 \Phi} = e^{-2 \Phi_0} (1-\rho^2) (1+r^2)~, \qquad C_2 = e^{-\Phi_0} (\rho^2 \extder \psi \wedge \extder \varphi - r^2 \extder t \wedge \extder \phi )~, \qquad C_4 = 0~.
\]
Further swapping
\[
r \leftrightarrow \rho~, \qquad \psi \leftrightarrow \phi~,
\]
in \eqref{eq:metric_maxdef} and \eqref{eq:backbi2_maxdef} gives the mirror model of the undeformed $\AdS_3 \times \Sp^3 \times \To^4$ superstring \cite{Arutyunov:2014jfa},
\[
\extder s^2 &= \frac{1}{1-r^2} (-\extder t ^2 + \extder r^2 ) + r^2 \extder \phi^2 + \frac{1}{1+\rho^2} (\extder \varphi^2 + \extder \rho^2) + \rho^2 \extder \psi^2 + \extder x^i \extder x^i~, \\
B &= \frac{r}{1-r^2} \extder t \wedge \extder r + \frac{\rho}{1+\rho^2} \extder \varphi \wedge \extder \rho~, \qquad e^{-2 \Phi} = e^{-2 \Phi_0} (1-r^2) (1+\rho^2)~, \\
C_2 &= -e^{-\Phi_0} (r^2 \extder \varphi \wedge \extder \phi + \rho^2 \extder t \wedge \extder \psi )~, \qquad C_4 = 0~.
\]
On the other hand, this is not the case for the other background \eqref{eq:backbi1_maxdef}.
\paragraph{Conclusions.}
Investigating limits of the two supergravity solutions for $\kappa_+=\kappa, \kappa_-=0$, which corresponds to the one-parameter $\eta$-deformation of the $\AdS_3 \times \Sp^3 \times \To^4$ superstring, we found that the background \eqref{eq:backbi1} has the expected Pohlmeyer limit, while the maximal deformation limit of the background \eqref{eq:backbi2} corresponds to the undeformed $\AdS_3 \times \Sp^3 \times \To^4$ mirror theory. We will now show that more is true: the background \eqref{eq:backbi2} actually exhibits mirror duality. 
\subsection{Mirror model and mirror duality}
The mirror model of the light-cone gauge-fixed string introduced in \cite{Arutyunov:2007tc} plays an important role in the Thermodynamic Bethe Ansatz approach and the calculation of finite-size corrections in the context of integrable models \cite{Ambjorn:2005wa}. It is constructed out of the original theory by performing a double Wick rotation in the worldsheet coordinates, $\tau \rightarrow i \sigma$, $\sigma \rightarrow i \tau$. While this transformation does not affect Lorentz-invariant theories (up to a parity reflection), this is no longer the case for a non-relativistic theory, whose mirror describes a new model. This is in particular true for the worldsheet theory of an $\AdS$ superstring upon light-cone gauge fixing, whose mirror defines a new two-dimensional quantum field theory. It is the thermodynamics of this new QFT that plays a central role in solving the spectral problem of $\AdS$/CFT using integrability. In addition to being a useful tool, one may wonder if there exists a more physical interpretation of this mirror model. This question was investigated in \cite{Arutyunov:2014cra,Arutyunov:2014jfa}, where it was shown that the latter can be seen as the light-cone gauge theory of a free string on a different, mirror, background. Furthermore, for the $\eta$-deformed $\AdS_n \times \Sp^n \times \To^{10-2n}$ superstring, $n=2,3,5$, it was observed that at the bosonic level the mirror background can also be reached directly from the original background by a field and parameter redefinition \cite{Arutyunov:2014cra}. This is the concept of mirror duality. In this section we construct the mirror background of \eqref{eq:backbi2} and show that the mirror duality also extends to the full background, including the dilaton and the R-R fluxes.
\paragraph{Mirror model.} We start by constructing the mirror model of the background \eqref{eq:backbi2}, first discussing the metric. If we denote the two light cone directions by $t$ and $\varphi$ and assume that there is no cross term $G_{t \varphi}=0$ in the metric, then the mirror metric is obtained by interchanging $G_{tt}$ and $1/G_{\varphi \varphi}$. \footnote{Alternatively, this can be seen as doing a T-duality in $t$ and $\varphi$ and then exchanging the two coordinates. This point of view has the advantage that the transformation of the dilaton under T-duality is known.} 
For the particular example of the deformed $\AdS_3 \times \Sp^3 \times \To^4$ superstring with $\kappa_+=\kappa$, $\kappa_- = 0$, the mirror metric is
\[
\label{eq:metric_mirror}
\extder s^2 &= -\frac{1+\kappa^2 r^2}{1-r^2}  \extder t^2 + \frac{\extder \rho^2}{(1-\kappa^2 \rho^2) (1+\rho^2)} + \rho^2 \extder \psi^2 \\
&\qquad + \frac{1-\kappa^2 \rho^2}{1+\rho^2} \extder \varphi^2 + \frac{\extder r^2}{(1+\kappa^2 r^2) (1-r^2)} + r^2 \extder \phi^2 + \extder x^i \extder x^i~.
\]
In principle the B-field also transforms but since it is a closed two-form we will drop it altogether. The mirror fluxes (tilded quantities) are related to the original fluxes through \cite{Arutyunov:2014jfa}
\[
\tilde{\RRF}_{a_1 \dots a_n} = i^{n}  \RRF_{a_1 \dots a_n} ~,
\] 
where $a_1 \dots a_n$ are flat indices and we choose the labelling of the transverse space to be common to the theory and its mirror. In order to disentangle the contributions from the dilaton and the R-R fluxes we solve the supergravity equations of motion. Applying this transformation rule to the background \eqref{eq:backbi2} with $\kappa_+=\kappa$, $\kappa_-=0$ yields the following potentials of the mirror model
\[
\label{eq:fluxes_mirror2}
e^{-2 \Phi}&=e^{-2 \Phi_0} \frac{(1-r^2)(1+\rho^2)}{P(\rho,r)^2}~, \qquad
P(\rho,r)=1-\kappa^2 \rho^2 r^2 ~,\\
C_2 &= \frac{e^{-\Phi_0}}{\sqrt{1+\kappa^2}P(\rho,r)} \big(\kappa^2 r^2 (1+\rho^2) \extder t \wedge \extder \phi -\kappa^2 \rho^2 (1-r^2) \extder \psi \wedge  \extder \varphi \\
& \qquad - (1- r^2) \extder \varphi \wedge \extder \phi + (1+\rho^2) \extder t \wedge \extder \psi \big) \\
C_4 &=- \frac{\kappa e^{-\Phi_0}}{\sqrt{1+\kappa^2}P(\rho,r)}   \big(r^2 (1+\rho^2) \extder t \wedge \extder \phi - \rho^2 (1-r^2 ) \extder \psi \wedge  \extder \varphi \\
& \qquad + (1- r^2) \extder \varphi \wedge \extder \phi -  (1+\rho^2) \extder t \wedge \extder \psi \big) \wedge J_2~.
\]
One immediately sees that the limit $\kappa \rightarrow 0$ gives the mirror model of undeformed $\AdS_3 \times \Sp^3 \times \To^4$, as expected.

\paragraph{Mirror duality.}
Now that we have constructed the mirror model, we can prove that the mirror duality extends to the full background if one stays within the realm of deformations generated by the R-matrix $\mathcal R_2$. Starting from the supergravity background \eqref{eq:backbi2} with $\kappa_+=\kappa$, $\kappa_-=0$, and rescaling 
\[
\label{eq:rescaling_kappa}
t \rightarrow \frac{t}{\kappa}~, \qquad \rho \rightarrow \frac{\rho}{\kappa}~, \qquad \varphi \rightarrow \frac{\varphi}{\kappa}~, \qquad r \rightarrow \frac{r}{\kappa}~,
\]
together with $x^i \rightarrow x^i /\kappa$ in the torus directions, defining $\hat{\kappa}=1/\kappa$ and then interchanging the coordinates
\[
\rho \leftrightarrow r~, \qquad \psi \leftrightarrow \phi~,
\] 
indeed gives the mirror background defined through \eqref{eq:metric_mirror} and \eqref{eq:fluxes_mirror2}, up to identifying $\hat{\kappa}$ and $\kappa$. \footnote{We also exploit the gauge symmetries of the potentials $C_2$ and $C_4$ to match the expressions.}




\section{Examples of generalised supergravity backgrounds for the two-parameter deformation of the \texorpdfstring{$\AdS_3 \times \Sp^3 \times \To^4$}{AdS3 x S3 x T4} superstring} 
\label{sec:generalised}
For completeness and to make the link with the literature, let us also extract the fluxes corresponding to deformations governed by non-unimodular R-matrices. For the $\alg{psu}(1,1|2)$ superalgebra, R-matrices associated to the Dynkin diagrams $\Circle - \otimes -\Circle$ or $\otimes - \Circle -\otimes$ are non-unimodular. Thus, for the  $\AdS_3 \times \Sp^3 \times \To^4$ superstrings, deformations based on R-matrices corresponding to the following Dynkin diagrams 

\[
\Circle - \otimes -\Circle \qquad \Circle - \otimes -\Circle   \\
\otimes - \Circle -\otimes \qquad \otimes - \Circle -\otimes   \\
\Circle - \otimes -\Circle \qquad \otimes - \Circle -\otimes   \\
\Circle - \otimes -\Circle \qquad \otimes - \otimes -\otimes  \\
\otimes - \Circle -\otimes \qquad \otimes - \otimes -\otimes 
\]
are expected to lead to generalised supergravity backgrounds. In the first two cases the $\alg{psu}(1,1|2)$ Dynkin diagram is the same in both copies of the algebras, while in the remaining three cases the Dynkin diagrams are different in the two copies.
\subsection{Dynkin diagram \texorpdfstring{$(\Circle - \otimes -\Circle)^2$}{o-x-o} }
We start by considering R-matrices associated to the Dynkin diagram 
\[
\label{eq:Dynkinoxooxo}
\Circle - \otimes -\Circle \qquad \Circle - \otimes -\Circle ~.
\]
In other words, when constructing $\mathcal R = (R_L, R_R)$, both $R_L$ and $R_R$ are R-matrices associated to the distinguished Dynkin diagram $\Circle - \otimes -\Circle$ of $\alg{psu}(1,1|2)$. As before, we keep the same action on the $\alg{su}(1,1)$ and $\alg{su}(2)$ algebras and hence we are left with two such R-matrices, namely $R_0$ given in \eqref{eq:R0} and 
\[
R_0'(M)_{ij} = -i  \epsilon_{ij} M_{ij}~, \qquad \epsilon = \begin{pmatrix}
0 & +1 & - 1 & -1 \\
-1 & 0 & -1 & - 1 \\
+ 1 & +1 & 0 & +1 \\
+1 & + 1 & -1 & 0
\end{pmatrix}~.
\]
We can then define the two following R-matrices governing the deformation of the $\AdS_3 \times \Sp^3$ supercoset,
\[
\mathcal{R}_0=\diag(R_0,-R_0)~, \qquad \mathcal{R}_0'=\diag(R_0,-R_0')~.
\]
To extract the fluxes we use the same parametrisation \eqref{eq:param}. The background corresponding to the R-matrix $\mathcal R_0 = (R_0, -R_0)$ is then \footnote{In our conventions the Hodge star $\star$ acts on a $n$-form $A_n = \frac{1}{n!} A_{\mu_1 \dots \mu_n} \extder X^{\mu_1} \wedge \dots \wedge \extder X^{\mu_n}$ as
	\[
	(\star A_n)_{\mu_1 \dots \mu_{d-n}} = \frac{1}{n!} \sqrt{-G} \epsilon_{\mu_1 \dots \mu_{d-n} \nu_1 \dots \nu_n} A^{\nu_1 \dots \nu_n}~,
	\]
	where $G$ is the determinant of the metric.
	The self-duality condition for the five-form reads $\RRF_5 = \star \RRF_5$.
}
\[
\label{eq:backoxooxo1}
\mathcal F_1 &=  N \hat{\mathcal F}_1 ~, \\
\mathcal F_3 &=  N \left(\hat{\mathcal F}_3 + \frac{2 \kappa_-}{1-\kappa_-^2} \hat{\mathcal F}_1 \wedge J_2\right) ~,\\
\mathcal F_5 &= N \left( (1+\star) \hat{\mathcal F}_1 \wedge J_2 \wedge J_2+ \frac{2 \kappa_-}{1-\kappa_-^2} \hat{\mathcal F}_3 \wedge J_2 \right)  ~, \\
N &= 2 \sqrt{\frac{1+\kappa_+^2}{1+\kappa_-^2}} \frac{1-\kappa_-^2}{\sqrt{F(\rho)\tilde{F}(r)}}~,
\]
were we introduced the auxiliary one-form and three-form
\[
\label{eq:backxoxxox1aux}
\hat{\mathcal F}_1 &= \kappa_- \left[ (1+\rho^2) \extder t + (1-r^2) \extder \varphi \right] + \kappa_+ \left[-\rho^2 \extder \psi + r^2 \extder \phi \right] ~, \\
\hat{\mathcal F}_3 &= \frac{1}{F(\rho)} \big[ \rho \, \extder t \wedge \extder \rho \wedge \extder \psi - \kappa_+^2 \rho r^2 \,\extder t \wedge \extder \rho \wedge \extder \phi - \kappa_-^2 \rho (1-r^2) \,\extder \rho \wedge \extder \psi \wedge \extder \varphi \\
&\qquad- \kappa_+ \kappa_- \rho (1-r^2) \,\extder t \wedge \extder \rho \wedge \extder \varphi - \kappa_+ \kappa_- \rho r^2 \,\extder \rho \wedge \extder \psi \wedge \extder \phi \big] \\
& + \frac{1}{\tilde{F}(r)} \big[ r \, \extder \varphi \wedge \extder r \wedge \extder \phi + \kappa_+^2 \rho^2 r \, \extder \psi \wedge \extder \varphi \wedge \extder r - \kappa_-^2 (1+\rho^2) r \, \extder t \wedge \extder r \wedge \extder \phi \\
&\qquad - \kappa_+ \kappa_- (1+\rho^2) r \, \extder t \wedge \extder \varphi \wedge \extder r + \kappa_+ \kappa_- \rho^2 r \, \extder \psi \wedge \extder r \wedge \extder \phi \big] ~. 
\]
Setting $\kappa_-=0$ and changing the sign of $\kappa_+$ gives the ABF-type background of \cite{Arutyunov:2015mqj}, obtained by starting from a supergravity solution with a dilaton linear in some isometries and formally dualising the metric and the fluxes in those isometries. The resulting background solves the generalised supergravity equations of motion \cite{Arutyunov:2015mqj,Wulff:2016tju}. These equations depend on a vector $X$, which can be split into a background Killing vector $I$, and a remaining part $Z$. The standard supergravity equations of motion correspond to $I=0$ and $Z=\extder \Phi$. The full background \eqref{eq:backoxooxo1} satisfies the generalised supergravity equations of motion with 
\[
I &= 2 \kappa_+(1+\kappa_-^2) \left(\frac{1+\rho^2}{F(\rho)} \extder t + \frac{1-r^2}{\tilde{F}(r)} \extder \varphi \right) + 2 \kappa_-  (1+\kappa_+^2) \left( -\frac{\rho^2}{F(\rho)} \extder \psi + \frac{r^2}{\tilde{F}(r)} \extder \phi\right)~, \\
Z &= \extder \left(\frac{1}{2} \log F(\rho) + \frac{1}{2} \log \tilde{F}(r)\right)~.
\]

While the auxiliary one-form $\hat{\mathcal F}_1$ and three-form $\hat{\mathcal F}_3$ are both invariant under the transformations \eqref{eq:transfo2}, this is not the case of the background \eqref{eq:backoxooxo1}, since the deformations parameters $\kappa_\pm$ do not appear on an equal footing in the fluxes. Similarly to what we observed for the two supergravity solutions in \secref{sec:sugrabackgrounds}, we find that applying the transformations \eqref{eq:transfo2} to \eqref{eq:backoxooxo1} yields a new background, which actually corresponds to choosing the R-matrix $\mathcal R_0' = (R_0, -R_0')$. It is also a generalised supergravity solution and has fluxes
\[
\label{eq:backoxooxo2}
\mathcal F_1 &=  N \hat{\mathcal F}_1 ~, \\
\mathcal F_3 &=  N \left(\hat{\mathcal F}_3 + \frac{2 \kappa_+}{1-\kappa_+^2} \hat{\mathcal F}_1 \wedge J_2\right) ~,\\
\mathcal F_5 &= N \left( (1+\star) \hat{\mathcal F}_1 \wedge J_2 \wedge J_2+ \frac{2 \kappa_+}{1-\kappa_+^2} \hat{\mathcal F}_3 \wedge J_2 \right)  ~, \\
N &= 2 \sqrt{\frac{1+\kappa_-^2}{1+\kappa_+^2}} \frac{1-\kappa_+^2}{\sqrt{F(\rho)\tilde{F}(r)}}~.
\]
On the other hand, contrary to the supergravity case, the transformations \eqref{eq:transfo1} do not leave the backgrounds invariant. Rather, they give two new generalised supergravity solutions. It is not clear if these correspond to deformations based on Drinfel'd Jimbo R-matrices. 
\paragraph{Relating the two backgrounds by TsT transformations.}
Interestingly, one can also go from one background to the other by doing TsT transformations on the torus that depend on the deformation parameters. This is a new feature that was not true for the supergravity backgrounds in \secref{sec:sugrabackgrounds}. Indeed, these had different dilatons and since a TsT transformation on the torus does not affect the dilaton, it cannot be sufficient to go from one background to the other. 

Starting from \eqref{eq:backoxooxo1} and dualising along one torus direction, say $x_1$, and then doing a metric and B-field preserving $\grp{SO}(2)$ rotation
\[
x_1 = \cos \theta y_1 - \sin \theta y_2~, \qquad x_2 = \sin \theta y_1 + \cos \theta y_2~, \qquad \theta = \arcsin \frac{\kappa_-}{\sqrt{1+\kappa_-^2}}~,
\]
and finally T-dualising in the new coordinate $y_1$ gives the intermediate background \footnote{For notational convenience we identify the $x$ and $y$ coordinates. In particular the Kähler form on the torus is $J_2 = \extder y_1 \wedge \extder y_2 - \extder x_3 \wedge \extder x_4 = \extder x_1 \wedge \extder x_2 - \extder x_3 \wedge \extder x_4$.}
\[
\label{eq:backoxooxoTsT}
\mathcal F_1 &=  N \hat{\mathcal F}_1 ~, \qquad 
\mathcal F_3 =  N \hat{\mathcal F}_3 ~,\qquad
\mathcal F_5 = N  (1+\star) \hat{\mathcal F}_1 \wedge J_2 \wedge J_2~, \\
 N &= 2 \frac{\sqrt{(1+\kappa_-^2)(1+\kappa_+^2)}}{\sqrt{F(\rho)\tilde{F}(r)}}~.
\]
The $\kappa_-=0$ point of \eqref{eq:backoxooxo1} is rotated by $\theta=0$ and is thus unaffected by this TsT transformation. Indeed, \eqref{eq:backoxooxoTsT} coincides with \eqref{eq:backoxooxo1} for $\kappa_-=0$. Starting from the intermediate background \eqref{eq:backoxooxoTsT} and doing the same steps, but now with rotation angle 
\[
\theta = \arcsin \frac{\kappa_+}{\sqrt{1+\kappa_+^2}} ~,
\]
we arrive at the background \eqref{eq:backoxooxo2} associated to the second R-matrix $\mathcal R_0' =(R_0,-R_0')$. This second TsT now leaves the $\kappa_+=0$ point of \eqref{eq:backoxooxoTsT} invariant and indeed \eqref{eq:backoxooxoTsT} and \eqref{eq:backoxooxo2} coincide when $\kappa_+=0$. The intermediary background \eqref{eq:backoxooxoTsT} thus interpolates between the $\kappa_-=0$ point of \eqref{eq:backoxooxo1} and the $\kappa_+=0$ point of \eqref{eq:backoxooxo2}.

\subsection{Dynkin diagram \texorpdfstring{$(\otimes-\Circle-\otimes)^2$}{x-o-x} }
Let us now consider R-matrices associated with the Dynkin diagram 
\[
\label{eq:Dynkinxoxxox}
\otimes-\Circle-\otimes\qquad \otimes-\Circle-\otimes~,
\]
so when constructing $\mathcal R = (R_L, R_R)$, both $R_L$ and $R_R$ are R-matrices associated to the Dynkin diagram $\otimes - \Circle -\otimes$ of $\alg{psu}(1,1|2)$.
Again, we fix the action on the $\alg{su}(1,1)$ and $\alg{su}(2)$ algebras to be the same as for the reference R-matrix $R_0$ and hence we are left with two such R-matrices, namely
\[
\mathds{P}_3 = \begin{pmatrix}
1 & 2 & 3 & 4 \\
1 & 3 & 4 & 2
\end{pmatrix}~, \qquad R_{\mathds{P}_3}(M)_{ij} = -i  \epsilon_{ij} M_{ij}~, \qquad \epsilon = \begin{pmatrix}
0 & +1 & + 1 & +1 \\
-1 & 0 & -1 & - 1 \\
- 1 & +1 & 0 & +1 \\
-1 & + 1 & -1 & 0
\end{pmatrix}~, \\
\mathds{P}_4 = \begin{pmatrix}
1 & 2 & 3 & 4 \\
3 & 1 & 2 & 4
\end{pmatrix}~, \qquad R_{\mathds{P}_4}(M)_{ij} = -i  \epsilon_{ij} M_{ij}~, \qquad \epsilon = \begin{pmatrix}
0 & +1 & - 1 & +1 \\
-1 & 0 & -1 & + 1 \\
+ 1 & +1 & 0 & +1 \\
-1 & - 1 & -1 & 0
\end{pmatrix}~.
\]
For $ R_{\mathds{P}_1}$ the central bosonic node corresponds to an element of $\alg{su}(2)$, while for $ R_{\mathds{P}_2}$ the central bosonic node corresponds to an element of $\alg{su}(1,1)$. Constructing the backgrounds corresponding to the two choices
\[
\mathcal{R}_3&=\diag(R_{\mathds{P}_3},-R_{\mathds{P}_3})~, &\qquad \mathcal{R}_4&=\diag(R_{\mathds{P}_3},-R_{\mathds{P}_4})~,
\]
for the first R-matrix $\mathcal R_3$ we find that the R-R fluxes are the same as \eqref{eq:backoxooxoTsT} up to a change of sign in  $t$ and $\psi$ and thus is also a generalised supergravity background. Furthermore, it is invariant under the transformations \eqref{eq:transfo2}. It is not, however, invariant under the redefinitions of \eqref{eq:transfo1}, which generate a new generalised supergravity solution. The latter is nothing else than the background associated to the R-matrix $\mathcal R_4$.
 
\begin{figure} \centering
	\includegraphics[scale=1]{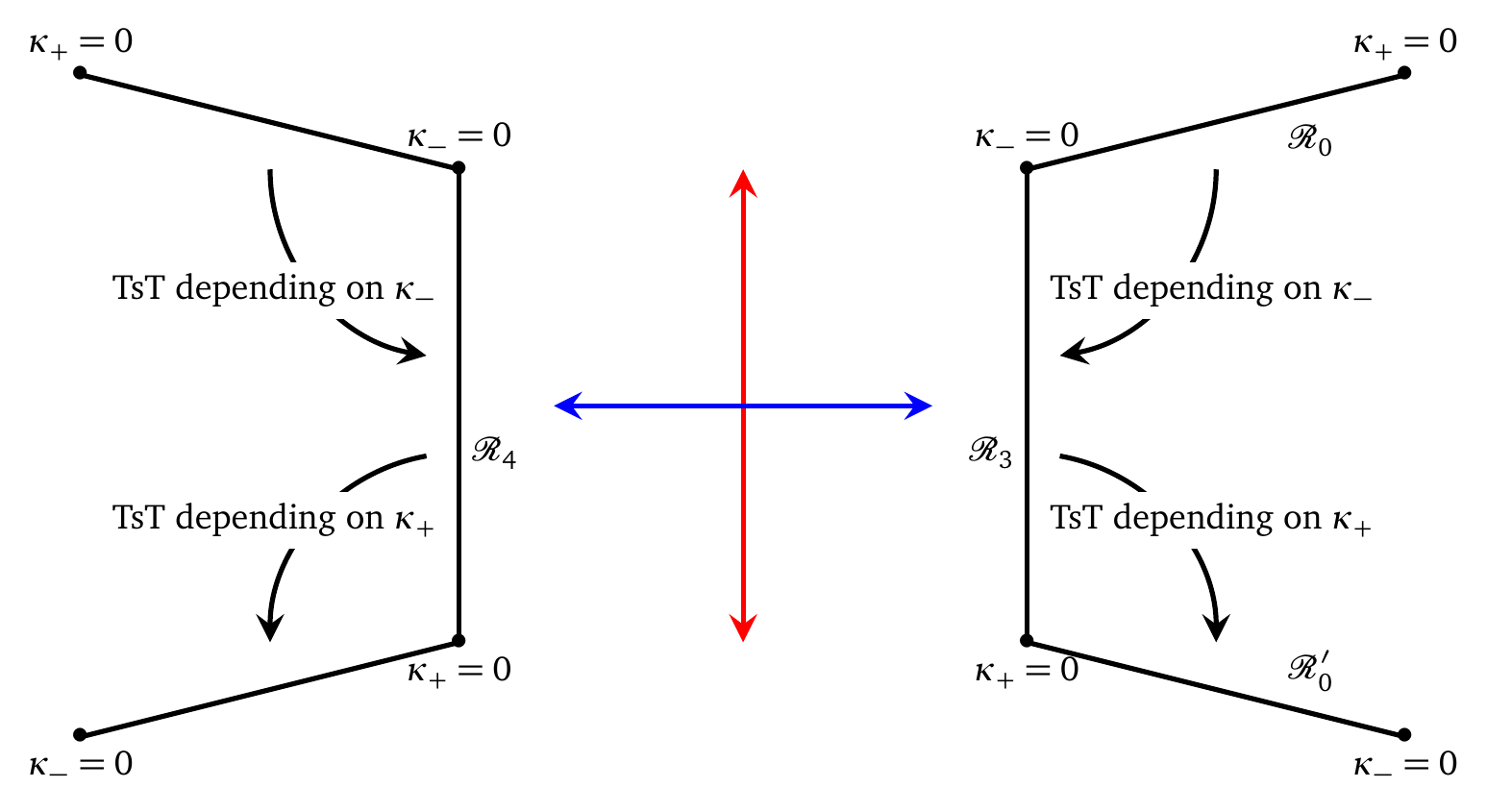}
	\caption{A window into the space of generalised supergravity backgrounds for the two-parameter deformation of the $\AdS_3 \times \Sp^3 \times \To^4$ superstring. The two vertical lines in the middle represent the two backgrounds based on R-matrices associated to the Dynkin diagram $(\otimes - \Circle - \otimes)^2$, namely $\mathcal R_3$ and $\mathcal R_4$. The top right and bottom right tilted lines correspond to the two backgrounds based on R-matrices associated to the Dynkin diagram $(\Circle - \otimes - \Circle)^2$, namely $\mathcal R_0$ and $\mathcal R_0'$. The transformation \protect\eqref{eq:transfo1} (respectively \protect\eqref{eq:transfo2}) is represented by the blue (respectively red) arrow. \protect\eqref{eq:transfo2} leaves the two backgrounds corresponding to $\mathcal R_3$ and $\mathcal R_4$ invariant and swaps the two backgrounds corresponding to $\mathcal R_0$ and $\mathcal R_0'$. The transformation \protect\eqref{eq:transfo1} swaps the backgrounds corresponding to $\mathcal R_3$ and $\mathcal R_4$ but does not leave the backgrounds associated to $\mathcal R_0$ or $\mathcal R_0'$ invariant. Rather it generates new generalised supergravity backgrounds (the two tilted lines on the left). It is not clear if these also correspond to particular Drinfel'd Jimbo R-matrices. Of course, since they are obtained from the $\mathcal R_0$ and $\mathcal R_0'$ backgrounds by a field redefinition they are also related to the $\mathcal R_4$ background by TsT transformations on the torus. }
	\label{fig:backgrounds_dist}
\end{figure}

\Figref{fig:backgrounds_dist} is a diagrammatic representation of the relations between the various generalised supergravity backgrounds that we have constructed.

\paragraph{Comparison with the literature.}
A proposal for the background of the two-parameter deformed $\AdS_3 \times \Sp^3 \times \To^4$ superstring has been derived in \cite{Araujo:2018rbc} by using a descent procedure involving the Page forms. It solves the generalised supergravity equations of motion and, up to signs, agrees with \eqref{eq:backoxooxoTsT}. We thus conclude that the background of \cite{Araujo:2018rbc} corresponds to the choice \eqref{eq:Dynkinxoxxox}.

\subsection{Limits} The generalised supergravity backgrounds associated to the Dynkin diagrams \eqref{eq:Dynkinoxooxo} and \eqref{eq:Dynkinxoxxox} are all related to the background \eqref{eq:backoxooxoTsT}, either via field redefinitions and/or T-dualities on the torus (see \figref{fig:backgrounds_dist}). Therefore we shall only discuss limits of the background \eqref{eq:backoxooxoTsT} here. First of all, implementing the transformations \eqref{eq:transfo_ppwave} and taking the limit $L \rightarrow \infty$ we find the same pp-wave background as for the two supergravity solutions. Second, the Pohlmeyer limit \eqref{eq:transfo_Pohlmeyer} is identical to \eqref{eq:Pohlmeyer}. Therefore in these two limits the generalised supergravity background actually becomes a standard supergravity background. On the other hand, it remains a generalised supergravity background in the maximal deformation limit and in particular does not match the mirror model. This behaviour is reminiscent of the $\eta$-deformed $\AdS_2 \times \Sp^2 \times \To^6$ superstring \cite{Hoare:2018ngg}.

\subsection{Different Dynkin diagrams in the two copies}
Let us finish this section by commenting on deformations constructed out of R-matrices associated to different Dynkin diagrams in the two copies of the $\alg{psu}(1,1|2)$ superalgebra. We take $\mathcal R=\diag(R_L, R_R)$ with $R_L$ associated to the Dynkin diagram $D_L$ and $R_R$ associated to the Dynkin diagram $D_R$.

When the deformation parameter in the right copy is set to zero, $\eta_R=0$ or equivalently $\kappa_+=\kappa_-$ then the choice of R-matrix in the right copy is not relevant since it is multiplied by $\eta_R=0$ in the action. Therefore at $\kappa_+=\kappa_-$ all R-matrices of the form $\mathcal R= \diag(R_L, -)$ (the symbol $-$ denoting any R-matrix) will give rise to the same background.  On the other hand, when the deformation parameter in the left copy $\eta_L=0$, or equivalently $\kappa_+=-\kappa_-$, then it is the choice of R-matrix in the left copy that does not play any role. Hence at $\kappa_+=-\kappa_-$ all R-matrices of the form $\mathcal R= \diag(-, R_R)$ give the same background.

Let us illustrate this for the R-matrices associated to the Dynkin diagram 
\[
\Circle - \otimes -\Circle \qquad \otimes - \Circle -\otimes~.
\]
The corresponding background will interpolate between the point $\kappa_+=\kappa_-$ of \eqref{eq:backoxooxo1} or \eqref{eq:backoxooxo2} (depending on the specific choice of Cartan-Weyl basis in the left copy of the algebra) and the point $\kappa_+=-\kappa_-$ of \eqref{eq:backoxooxoTsT} or \eqref{eq:backoxooxoTsT} followed by the field redefinition \eqref{eq:transfo1} (depending on the specific choice of Cartan-Weyl basis in the right copy of the algebra).

Also R-matrices associated to the Dynkin diagrams
\[
\label{eq:Dynkinoxoxxx}
\Circle - \otimes - \Circle \qquad \otimes - \otimes - \otimes
\]
or
 \[
 \label{eq:Dynkinxoxxxx}
 \otimes - \Circle - \otimes \qquad \otimes - \otimes - \otimes
 \]
are of this type. The novelty here is that the R-matrix is non-unimodular in one copy of the $\alg{psu}(1,1|2)$ superalgra, but unimodular in the other copy. Let us focus on the first Dynkin diagram \eqref{eq:Dynkinoxoxxx}, the analysis being similar for the other choice. At the point $\kappa_+=\kappa_-$, depending on the specific choice of Cartan-Weyl basis in the left copy of the algebra, the background will then coincide either with \eqref{eq:backoxooxo1} or \eqref{eq:backoxooxo2}.  In particular, it will still be a generalised supergravity background. On the other hand, when $\kappa_+=-\kappa_-$, the background will then become the supergravity solution \eqref{eq:backbi1} or \eqref{eq:backbi2}. This provides an example of a two-parameter integrable deformation that interpolates between a generalised supergravity and a standard supergravity solution. 

\section{Conclusions}
\label{sec:conclusions}

In this paper we proposed an explicit formula for the R-R fluxes of the two-parameter deformation of the Metsaev-Tseytlin action for supercosets with isometry group of the form $\hat{\grp{G}} \times \hat{\grp{G}}$. We then applied this result to the $\AdS_3 \times \Sp^3 \times \To^4$ superstring and constructed two supergravity backgrounds analogous to the $a=0$ and $a=1$  solutions of \cite{Lunin:2014tsa}. The two solutions correspond to two different choices of Drinfel'd Jimbo R-matrices satisfying the non-split modified classical Yang-Baxter equation on the $\alg{psu}(1,1|2)_L \oplus \alg{psu}(1,1|2)_R$ superisometry algebra. The two R-matrices both correspond to the fully fermionic Dynkin diagram 
\[
\otimes - \otimes - \otimes  \qquad \otimes - \otimes - \otimes~,
\]
and satisfy the unimodularity property of \cite{Borsato:2016ose}. They differ, however, in the specific choice of Cartan-Weyl basis used in the two copies of the $\alg{psu}(1,1|2)$ superalgebra.

The two solutions have the same metric and B-field, but different dilatons and three-form and five-form R-R fluxes. This provides further new examples of different embeddings of a given bosonic background into supergravity. The two solutions are related by a complex field redefinition and swapping of the deformation parameters. It would be interesting to understand if they can also be related by a set of target space dualities, for instance fermionic T-dualities. Such a transformation should modify the dilaton but leave the metric invariant. 

Let us stress that the existence of two supergravity backgrounds is not particular to the two-parameter deformation -- this property is still present after setting $\eta_L = \eta_R$, which corresponds to the usual one-parameter $\eta$-deformed $\AdS_3 \times \Sp^3 \times \To^4$ superstring. Studying limits of the two $\eta$-deformed $\AdS_3 \times \Sp^3 \times \To^4$ supergravity backgrounds we found that they have different Pohlmeyer and maximal deformation limits. Taking the Pohlmeyer limit of one of the solutions gives a pp-wave supergravity background whose light-cone gauge-fixing gives the Pohlmeyer reduced theory of the $\AdS_3 \times \Sp^3 \times \To^4$ superstring. On the other hand, the other background is not finite in that limit. For the maximal deformation limit, both backgrounds are finite but only the latter solution gives the mirror $\AdS_3 \times \Sp^3 \times \To^4$ background. Motivated by this observation we constructed its mirror theory and showed that the mirror duality previously observed in \cite{Arutyunov:2014cra} also extends to the full supergravity background in that case. 

Another interesting limit of the supergravity backgrounds is the type of homogeneous limit considered in \cite{Hoare:2016hwh,vanTongeren:2019dlq}. By taking particular singular boosts of the $\eta$-deformed backgrounds one can construct homogeneous Yang-Baxter deformations \cite{Kawaguchi:2014qwa}, based on R-matrices solving the classical Yang-Baxter equation. This boosting procedure has recently been applied to find new homogeneous R-matrices involving odd generators and construct unimodular jordanian R-matrices for $\alg{psu}(2,2|4)$ \cite{vanTongeren:2019dlq}. It would be interesting to explore this type of limit for the two inhomogeneous R-matrices and associated supergravity background considered in this paper.

To complete our study we also explicitly constructed generalised supergravity backgrounds associated to the Dynkin diagrams
\[
\Circle - \otimes - \Circle  \qquad \Circle - \otimes - \Circle ~,
\]
and
\[
\label{eq:xoxxox}
\otimes - \Circle - \otimes  \qquad \otimes - \Circle - \otimes~.
\]
The various generalised supergravity backgrounds are all related by (complex) field redefinitions and/or T-dualities on the torus. Moreover, we were able to make the link with the results of \cite{Araujo:2018rbc} obtained by a descent procedure involving the page forms, which appear to correspond to the choice \eqref{eq:xoxxox}. It remains to be understood why this is the case. 

\medskip

Can these results be used to answer some of the questions raised in the context of $\eta$-deformations of other-dimensional spaces? In the  $\AdS_2 \times \Sp^2 \times \To^6$ case, a family of supergravity solutions with parameter $a$ interpolating between $0$ and $1$ has been constructed in \cite{Lunin:2014tsa}. The background at $a=0$ has explicit mirror duality. However, in \cite{Hoare:2018ngg} only one supergravity background of the $\eta$-deformed $\AdS_2 \times \Sp^2 \times \To^6$ superstring was found, matching the $a=1$ solution. It remains unclear how the other solutions $a \in [0,1)$ can be generated and if there exists an R-matrix giving rise to the interpolating background, within or outside the realm of Drinfel'd Jimbo R-matrices. To gain more insight it might be interesting to study consistent truncations of the different $\AdS_3 \times \Sp^3 \times \To^4$ backgrounds down to $4$ dimensions. For the $\eta$-deformed $\AdS_5 \times \Sp^5$ superstring, finding a supergravity background with explicit mirror duality remains an important open question.

Furthermore, the fact that the perturbative worldsheet S-matrix obtained in \cite{Arutyunov:2015qva} for the $\AdS_5\times \Sp^5$ superstring does not match the expansion of the exact light-cone gauge fixed S-matrix of \cite{Beisert:2008tw} when including fermions remains a puzzling problem. Focussing on $\AdS_3$ one could calculate the light-cone gauge fixed S-matrices describing scattering above BMN strings \cite{Berenstein:2002jq} for the two supergravity solutions and compare them to each other. Deformed R-matrices have been constructed in \cite{Hoare:2014oua} but the overall dressing phases obeying unitary, braiding unitary and crossing symmetry remains to be found.

The presence of a naked singularity for some value of $\rho$ due to the non-compactness of the original $\AdS$ space is also a long standing problem in the context of $\eta$-deformations. While this singularity cannot be avoided for $\AdS_2$ and $\AdS_5$, the case of $\AdS_3$ is special and offers the intriguing possibility of obtaining a smooth deformation. Indeed, when one of the deformation parameters vanishes, either $\eta_L=0$ or $\eta_R=0$, then the curvature singularity disappears. The metric becomes the one of a warped $\AdS_3$ and squashed $\Sp^3$ metric, together with the flat $\To^4$ metric.  Moreover, half of the supersymmetries of the original theory are preserved. The two supergravity backgrounds \eqref{eq:backbi1} and \eqref{eq:backbi2} coincide, with constant dilaton. Backgrounds containing warped $\AdS_3$ or squashed $\Sp^3$ geometries are related to $\AdS_3 \times \Sp^3$ via T-duality \cite{Orlando:2010yh}. It would be interesting to analyse what happens upon including fermions.


\medskip

Another direction concerns the relation of the two-parameter deformation to other known integrable deformations. Indeed, the $\eta$-deformation is a Poisson-Lie symmetric model \cite{Klimcik:1995ux,Klimcik:1995jn} and it has been conjectured in \cite{Hoare:2015gda} that its Poisson-Lie dual in the full superisometry algebra is related via analytic continuation to the $\lambda$ model of \cite{Hollowood:2014qma}. The latter is yet another instance of an integrable $q$-deformation, but with $|q|=1$. In contrast to the $\eta$-deformation, the $\lambda$-deformation always defines a critical string theory \cite{Borsato:2016ose}. Embeddings of the $\lambda$-deformed $\AdS_3 \times \Sp^3$ metric into supergravity have been proposed in \cite{Sfetsos:2014cea} and \cite{Chervonyi:2016ajp}. It is an interesting open question to find the Poisson-Lie duals of the two-parameter deformations considered in this paper and compare them to the multi-parameter $\lambda$-deformation constructed in \cite{Georgiou:2018gpe}.

\medskip

Last but not least, for supercosets with superisometry algebra $\alg{g} \oplus \alg{g}$ a Wess-Zumino-Witten term can be added to the action, corresponding to the introduction of NS-NS flux \cite{Cagnazzo:2012se}.  Combining the WZW term with the two-parameter deformation, one can construct an integrable three-parameter deformation of the semi-symmetric space sigma model \cite{Delduc:2018xug}. It would be interesting to obtain a closed formula for the NS-NS and R-R fluxes and see if unimodular R-matrices also give rise to Weyl-invariant theories in this more general setting. Starting from the undeformed pure NS-NS background and requiring that the deformation does not turn on R-R fluxes, one could then investigate the possibility of finding marginal deformations of the WZW point (see \cite{Borsato:2018spz} for related work in the context of homogeneous Yang-Baxter deformations). This is particularly interesting because these new pure NS-NS backgrounds could then be analysed with integrability and CFT techniques, which have recently been used to investigate the dual CFTs of $\AdS_3$ superstrings with pure NS-NS fluxes \cite{Gaberdiel:2018rqv,Giribet:2018ada,Eberhardt:2018ouy,Eberhardt:2019qcl}. This would then open the door to the exciting possibility of finding CFT duals of the $\eta$-deformed theories.

\section*{Acknowledgements}

Part of this work has been presented at the \textit{Integrability, duality and beyond} workshop in Santiago de Compostela and I would like to thank the organisers and participants for interesting discussions. 
Many thanks also to B.~Hoare for guidance throughout this project, R.~Borsato, A.~Sfondrini, A.~Tseytlin and L.~Wulff for valuable discussions and B.~Hoare and S.~van Tongeren for helpful comments on the draft.
This work is supported by grant no.~615203 from the European Research Council under the FP7.

\appendix

\section{Conventions}
\label{app:conventions}
\subsection{Gamma matrices}
\label{app:Gamma}
\paragraph{Pauli matrices.}
The gamma matrices are constructed out of the $2 \times 2$ Pauli matrices
\[
\sigma_1 = \begin{pmatrix} 0 & 1 \\ 1 & 0 \end{pmatrix}~, \qquad \sigma_2 = \begin{pmatrix} 0 & -i \\ i & 0 \end{pmatrix}~, \qquad \sigma_3 = \begin{pmatrix} 1 & 0 \\ 0 & -1 \end{pmatrix}~.
\]
\paragraph{Gamma matrices.}
We choose the following basis for the ten $32 \times 32$ dimensional gamma matrices,
\[
\Gamma^0 &= -i \sigma_1 \otimes \identity \otimes \identity \otimes \sigma_3 \otimes \identity~, &\qquad  \Gamma^1 &= \sigma_1 \otimes \sigma_3 \otimes \sigma_2 \otimes \sigma_2 \otimes \identity~, \\
\Gamma^2 &= \sigma_1 \otimes \identity \otimes \identity \otimes \sigma_1 \otimes \identity~, &\qquad 
\Gamma^3 &= -\sigma_2 \otimes \identity \otimes \identity \otimes \identity \otimes \sigma_3~, \\
\Gamma^4 &= \sigma_2 \otimes \identity \otimes \sigma_2 \otimes \identity \otimes \sigma_2~, &\qquad  \Gamma^5 &= \sigma_2 \otimes \identity \otimes \identity \otimes \identity \otimes \sigma_1~, \\
\Gamma^6 &= \sigma_1 \otimes \sigma_2 \otimes \identity \otimes \sigma_2 \otimes \identity~, &\qquad
\Gamma^7 &= \sigma_1 \otimes \sigma_1 \otimes \sigma_2 \otimes \sigma_2 \otimes \identity \\
\Gamma^8 &= \sigma_2 \otimes \sigma_2 \otimes \sigma_1 \otimes \identity \otimes \sigma_2 ~, &\qquad \Gamma^9 &= \sigma_2 \otimes \sigma_2 \otimes \sigma_3 \otimes \identity \otimes \sigma_2~.
\]
They satisfy the Clifford algebra $\anticom{\Gamma^a}{\Gamma^b} = 2 \eta^{ab}$ with
\[
\Gamma^{11}= \Gamma^0 \Gamma^1 \cdots \Gamma^9 = \sigma_3 \otimes \identity_{16}~, \qquad \mathcal{C} = i \sigma_2 \otimes \identity_{16}~, \qquad (\mathcal{C} \Gamma^a)^t = \mathcal{C} \Gamma^a~.
\]
The ten $16 \times 16$ chiral blocks $\gamma^a$ are then identified using
\[
\Gamma^a = \begin{pmatrix}
0 & (\gamma^a)^{\alpha \beta} \\
(\gamma^a)_{\alpha \beta} & 0
\end{pmatrix}~.
\]
and satisfy $\gamma^a_{\alpha \beta} (\gamma^b)^{\beta \gamma} + \gamma^b_{\alpha \beta} (\gamma^a)^{\beta \gamma} =2 \eta^{ab} \delta_\alpha^\gamma $.
The projector
\def\projF{\text{Proj}}
\[
\projF = \frac{1}{2} (\identity_{16} + \gamma^{6789}) = \diag(1,0) \otimes \identity_8 ~,
\]
projects onto a $8$-dimensional spinor subspace and thus can be used to effectively make these matrices $8 \times 8$ with spinor index $\alpha=1,\dots,8$.
In particular, we have
\[
\projF \gamma^a \projF &\rightarrow \bar{\gamma}^a~, & \quad a & =0,\dots,5~, \\
\projF \gamma^a \projF &\rightarrow 0~, & \quad a & = 6,7,8,9~,
\]
where the arrow represents the projection onto $8 \times 8$ matrices.
\subsection{Generators of $\alg{psu}(1,1|2)$}
The real form $\alg{su}(1,1|2)$ is given by those elements of the complexified superalgebra $\alg{sl}(2|2;\Complex)$ satisfying
\[ \label{eq:realitycond} M^\dagger H + H M =0~, \qquad H=\begin{pmatrix} \sigma_3 & 0 \\ 0 & \identity_2 \end{pmatrix}~.
\]
The superalgebra $\alg{su}(1,1|2)$ contains the $1$-dimensional ideal $\alg{u}(1)$ generated by $i \identity_4$.
The quotient of $\alg{su}(1,1|2)$ over this $\alg{u}(1)$ subalgebra defines the superalgebra $\alg{psu}(1,1|2)$.

\paragraph{Bosonic generators.}
Our choice for the three $\alg{su}(1,1)$ generators is
\[
L_1 = \frac{1}{2} \begin{pmatrix} \sigma_1 & 0 \\ 0 & 0 \end{pmatrix} ~, \qquad
L_2 = \frac{1}{2} \begin{pmatrix} \sigma_2 & 0 \\ 0 & 0 \end{pmatrix} ~, \qquad
L_3 = \frac{1}{2} \begin{pmatrix} i\sigma_3 & 0 \\ 0 & 0 \end{pmatrix}~, \\
\]
and for the three $\alg{su}(2)$ generators is
\[
J_1 = \frac{1}{2} \begin{pmatrix} 0 & 0 \\ 0 & i\sigma_1 \end{pmatrix} ~, \qquad
J_2 = \frac{1}{2} \begin{pmatrix} 0 & 0 \\ 0 & -i \sigma_2 \end{pmatrix}~, \qquad
J_3 =\frac{1}{2} \begin{pmatrix} 0 & 0 \\ 0 & i \sigma_3 \end{pmatrix} ~.
\]
\paragraph{Fermionic generators.}
The $\alg{psu}(1,1|2)$ superalgebra also contains eight fermionic generators $Q_{\ind{I} \check{\alpha} \hat{\alpha}}$, where $I=1,2$, $\check{\alpha}=1,2$ is the  $\alg{su}(1,1)$ spinor index and $\hat{\alpha}=1,2$ is the $\alg{su}(2)$ spinor index.
We choose them to be
\[
Q_{1 \check{\alpha} \hat{\alpha}} = \frac{1}{\sqrt{2}} i^{(\check{\alpha}-\hat{\alpha})} \begin{pmatrix} 0 & N_{\check{\alpha} \hat{\alpha}} \\ i \sigma_3 (N_{\check{\alpha} \hat{\alpha}})^t \sigma_3 & 0 \end{pmatrix}~,
\\
Q_{2 \check{\alpha} \hat{\alpha}} = \frac{1}{\sqrt{2}} i^{(\check{\alpha}-\hat{\alpha})} \begin{pmatrix} 0 & i N_{\check{\alpha} \hat{\alpha}} \\ \sigma_3 (N_{\check{\alpha} \hat{\alpha}})^t \sigma_3 & 0 \end{pmatrix}~.
\]
The indices $I$, $\check{\alpha}$ and $\hat{\alpha}$ can be gathered into a single index, $\alpha=1,\dots,8$, and we define the generators $Q_{\alpha}$ as
\[
Q_{1} &= Q_{111}~, &\qquad Q_{2} &= Q_{112}~, &\qquad Q_{3} &= Q_{121}~, &\qquad Q_{4} &= Q_{122}~, \\
Q_{5} &= Q_{211}~, &\qquad Q_{6} &= Q_{212}~, &\qquad Q_{7} &= Q_{221}~, &\qquad Q_{8} &= Q_{222}~.
\]
While these generators themselves do not satisfy the reality condition \eqref{eq:realitycond}, elements of the Grassmann envelope $\theta^{\alpha} Q_{\alpha}$ will do so if one imposes suitable reality conditions on the fermions $\theta^{\alpha}$.

\subsection{Two copies of $\alg{psu}(1,1|2)$}
For elements in the Lie algebra $\alg{psu}(1,1|2)_L \oplus \alg{psu}(1,1|2)_R$ we use the standard block-diagonal matrix realisation $\mathcal X = \diag(X^L, X^R)$ with $X^L \in \alg{psu}(1,1|2)_L$ and $X^R \in \alg{psu}(1,1|2)_R$. Furthermore, the generators are chosen so that they belong to a specific $\Integer_4$ grading as defined in \eqref{eq:grading}. The elements of grade 0 generate the $\alg{su}(1,1) \oplus \alg{su}(2)$ subalgebra and are
\[
\mathcal  J_{01} &= \diag(L_2,L_2)~, &\qquad \mathcal  J_{02} &= -\diag(L_1,L_1)~, &\qquad \mathcal  J_{12} &= -\diag(L_3,L_3)~,  \\
\mathcal  J_{34} &= \diag(J_2,J_2)~, &\qquad \mathcal  J_{35} &= -\diag(J_1,J_1)~, &\qquad \mathcal  J_{45} &= \diag(J_3,J_3)~.
\]
The ones of grade 2 are given by
\[
\mathcal P_0 &= \diag(L_3,-L_3)~, &\qquad\mathcal P_1 &= \diag(L_1,-L_1)~, &\qquad\mathcal P_2 &= \diag(L_2,-L_2)~, \\
\mathcal P_3 &= \diag(J_3,-J_3)~, &\qquad\mathcal P_4 &= \diag(J_1,-J_1)~, &\qquad\mathcal P_5 &= \diag(J_2, -J_2)~.
\]
Finally, the fermionic generators of grade 1 are denoted by $\mathcal Q_{1 \alpha}$, while the ones of grade 3 are denotes by $\mathcal Q_{2 \alpha}$. They are defined through
\[
\mathcal Q_{1 \alpha} =  \diag(Q_{\alpha}, -i Q_{\alpha} )~, \qquad 
\mathcal Q_{2 \alpha} =  \diag(Q_{\alpha}, i Q_{\alpha} )~.
\]
Our choice of generators matches the conventions of \cite{Borsato:2016ose}, in particular we have the anti\hyp{}commutation relations
\[
\anticom{\mathcal Q_{1 \alpha}}{\mathcal Q_{1 \beta}}= \anticom{\mathcal Q_{2 \alpha}}{\mathcal Q_{2 \beta}} = i \bar{\gamma}^a_{\alpha \beta} \mathcal P_a~.
\]

\begin{bibtex}[\jobname]
	@article{Hoare:2018ngg,
		author         = "Hoare, Ben and Seibold, Fiona K.",
		title          = "{Supergravity backgrounds of the $\eta$-deformed AdS$_2
			\times S^2 \times T^6 $ and AdS$_5 \times S^5$
			superstrings}",
		journal        = "JHEP",
		volume         = "01",
		year           = "2019",
		pages          = "125",
		doi            = "10.1007/JHEP01(2019)125",
		eprint         = "1811.07841",
		archivePrefix  = "arXiv",
		primaryClass   = "hep-th",
		SLACcitation   = "
	}
@article{Maldacena:1997re,
	author         = "Maldacena, Juan Martin",
	title          = "{The Large N limit of superconformal field theories and
		supergravity}",
	journal        = "Int. J. Theor. Phys.",
	volume         = "38",
	year           = "1999",
	pages          = "1113-1133",
	doi            = "10.1023/A:1026654312961, 10.4310/ATMP.1998.v2.n2.a1",
	note           = "[Adv. Theor. Math. Phys.2,231(1998)]",
	eprint         = "hep-th/9711200",
	archivePrefix  = "arXiv",
	primaryClass   = "hep-th",
	reportNumber   = "HUTP-97-A097, HUTP-98-A097",
	SLACcitation   = "
}
@article{Witten:1998qj,
	author         = "Witten, Edward",
	title          = "{Anti-de Sitter space and holography}",
	journal        = "Adv. Theor. Math. Phys.",
	volume         = "2",
	year           = "1998",
	pages          = "253-291",
	doi            = "10.4310/ATMP.1998.v2.n2.a2",
	eprint         = "hep-th/9802150",
	archivePrefix  = "arXiv",
	primaryClass   = "hep-th",
	reportNumber   = "IASSNS-HEP-98-15",
	SLACcitation   = "
}
@article{Grigoriev:2008jq,
	author         = "Grigoriev, M. and Tseytlin, Arkady A.",
	title          = "{On reduced models for superstrings on $AdS_n \times S^n$}",
	booktitle      = "{Progress of string theory and quantum field theory.
		Proceedings, International Conference, Osaka, Japan,
		December 7-10, 2007}",
	journal        = "Int. J. Mod. Phys.",
	volume         = "A23",
	year           = "2008",
	pages          = "2107-2117",
	doi            = "10.1142/S0217751X08040652",
	eprint         = "0806.2623",
	archivePrefix  = "arXiv",
	primaryClass   = "hep-th",
	reportNumber   = "IMPERAL-TP-AT-2008-2",
	SLACcitation   = "
}
	@article{Metsaev:1998it,
		author         = "Metsaev, R. R. and Tseytlin, Arkady A.",
		title          = "{Type IIB superstring action in $AdS_5 \times S^5$ background}",
		journal        = "Nucl. Phys.",
		volume         = "B533",
		year           = "1998",
		pages          = "109-126",
		doi            = "10.1016/S0550-3213(98)00570-7",
		eprint         = "hep-th/9805028",
		archivePrefix  = "arXiv",
		primaryClass   = "hep-th",
		reportNumber   = "FIAN-TD-98-21, IMPERIAL-TP-97-98-44, NSF-ITP-98-055",
		SLACcitation   = "
	}
	
	@article{Berkovits:1999zq,
		author         = "Berkovits, N. and Bershadsky, M. and Hauer, T. and Zhukov, S. and Zwiebach, B.",
		title          = "{Superstring theory on $AdS_2 \times S^2$ as a coset supermanifold}",
		journal        = "Nucl. Phys.",
		volume         = "B567",
		year           = "2000",
		pages          = "61-86",
		doi            = "10.1016/S0550-3213(99)00683-5",
		eprint         = "hep-th/9907200",
		archivePrefix  = "arXiv",
		primaryClass   = "hep-th",
		reportNumber   = "IFT-P-060-99, HUTP-99-A044, MIT-CTP-2878, CTP-MIT-2878",
		SLACcitation   = "
	}
	
	@article{Borsato:2016ose,
		author         = "Borsato, Riccardo and Wulff, Linus",
		title          = "{Target space supergeometry of $\eta$ and $\lambda$-deformed strings}",
		journal        = "JHEP",
		volume         = "10",
		year           = "2016",
		pages          = "045",
		doi            = "10.1007/JHEP10(2016)045",
		eprint         = "1608.03570",
		archivePrefix  = "arXiv",
		primaryClass   = "hep-th",
		reportNumber   = "IMPERIAL-TP-LW-2016-03",
		SLACcitation   = "
	}
	
	@article{Arutyunov:2015mqj,
		author         = "Arutyunov, G. and Frolov, S. and Hoare, B. and Roiban, R. and Tseytlin, A. A.",
		title          = "{Scale invariance of the $\eta$-deformed $AdS_5 \times S^5$ superstring, T-duality and modified type II equations}",
		journal        = "Nucl. Phys.",
		volume         = "B903",
		year           = "2016",
		pages          = "262-303",
		doi            = "10.1016/j.nuclphysb.2015.12.012",
		eprint         = "1511.05795",
		archivePrefix  = "arXiv",
		primaryClass   = "hep-th",
		reportNumber   = "ZMP-HH-15-27, TCDMATH-15-12, IMPERIAL-TP-AT-2015-08",
		SLACcitation   = "
	}
	
	@article{Wulff:2016tju,
		author         = "Tseytlin, A. A. and Wulff, L.",
		title          = "{Kappa-symmetry of superstring sigma model and generalized 10d supergravity equations}",
		journal        = "JHEP",
		volume         = "06",
		year           = "2016",
		pages          = "174",
		doi            = "10.1007/JHEP06(2016)174",
		eprint         = "1605.04884",
		archivePrefix  = "arXiv",
		primaryClass   = "hep-th",
		reportNumber   = "IMPERIAL-TP-LW-2016-02",
		SLACcitation   = "
	}
	
	@article{Arutyunov:2013ega,
		author         = "Arutyunov, Gleb and Borsato, Riccardo and Frolov, Sergey",
		title          = "{S-matrix for strings on $\eta$-deformed $AdS_5 \times S^5$}",
		journal        = "JHEP",
		volume         = "04",
		year           = "2014",
		pages          = "002",
		doi            = "10.1007/JHEP04(2014)002",
		eprint         = "1312.3542",
		archivePrefix  = "arXiv",
		primaryClass   = "hep-th",
		reportNumber   = "ITP-UU-13-31, SPIN-13-23, HU-MATHEMATIK-2013-24, TCD-MATH-13-16",
		SLACcitation   = "
	}
	
	@article{Arutyunov:2015qva,
		author         = "Arutyunov, Gleb and Borsato, Riccardo and Frolov, Sergey",
		title          = "{Puzzles of $\eta$-deformed $AdS_5 \times S^5$}",
		journal        = "JHEP",
		volume         = "12",
		year           = "2015",
		pages          = "049",
		doi            = "10.1007/JHEP12(2015)049",
		eprint         = "1507.04239",
		archivePrefix  = "arXiv",
		primaryClass   = "hep-th",
		reportNumber   = "ITP-UU-15-10, TCD-MATH-15-05, ZMP-HH-15-19",
		SLACcitation   = "
	}
	
	@article{Arutynov:2014ota,
		author         = "Arutyunov, Gleb and de Leeuw, Marius and van Tongeren, Stijn J.",
		title          = "{The exact spectrum and mirror duality of the $(AdS_5 \times S^5)_\eta$ superstring}",
		journal        = "Theor. Math. Phys.",
		volume         = "182",
		year           = "2015",
		number         = "1",
		pages          = "23-51",
		doi            = "10.1007/s11232-015-0243-9",
		note           = "[\textsf{Teor. Mat. Fiz. 182, 28 (2014)}]",
		eprint         = "1403.6104",
		archivePrefix  = "arXiv",
		primaryClass   = "hep-th",
		SLACcitation   = "
	}
	
	@article{Klabbers:2017vtw,
		author         = "Klabbers, Rob and van Tongeren, Stijn J.",
		title          = "{Quantum Spectral Curve for the $\eta$-deformed $AdS_5 \times S^5$ superstring}",
		journal        = "Nucl. Phys.",
		volume         = "B925",
		year           = "2017",
		pages          = "252-318",
		doi            = "10.1016/j.nuclphysb.2017.10.005",
		eprint         = "1708.02894",
		archivePrefix  = "arXiv",
		primaryClass   = "hep-th",
		reportNumber   = "ZMP-HH-17-25, HU-EP-17-21",
		SLACcitation   = "
	}
	
	@article{Arutyunov:2014cra,
		author         = "Arutyunov, Gleb and van Tongeren, Stijn J.",
		title          = "{$AdS_5 \times S^5$ mirror model as a string sigma model}",
		journal        = "Phys. Rev. Lett.",
		volume         = "113",
		year           = "2014",
		pages          = "261605",
		doi            = "10.1103/PhysRevLett.113.261605",
		eprint         = "1406.2304",
		archivePrefix  = "arXiv",
		primaryClass   = "hep-th",
		reportNumber   = "HU-EP-14-21, HU-MATH-14-12, ITP-UU-14-18, SPIN-14-16",
		SLACcitation   = "
	}
	
	@article{Arutyunov:2014jfa,
		author         = "Arutyunov, Gleb and van Tongeren, Stijn J.",
		title          = "{Double Wick rotating Green-Schwarz strings}",
		journal        = "JHEP",
		volume         = "05",
		year           = "2015",
		pages          = "027",
		doi            = "10.1007/JHEP05(2015)027",
		eprint         = "1412.5137",
		archivePrefix  = "arXiv",
		primaryClass   = "hep-th",
		SLACcitation   = "
	}
	
	@article{Pachol:2015mfa,
		author         = "Pacho\l{}, Anna and van Tongeren, Stijn J.",
		title          = "{Quantum deformations of the flat space superstring}",
		journal        = "Phys. Rev.",
		volume         = "D93",
		year           = "2016",
		pages          = "026008",
		doi            = "10.1103/PhysRevD.93.026008",
		eprint         = "1510.02389",
		archivePrefix  = "arXiv",
		primaryClass   = "hep-th",
		reportNumber   = "HU-EP-15-48, HU-MATH-15-13",
		SLACcitation   = "
	}
	
	@article{Beisert:2008tw,
		author         = "Beisert, Niklas and Koroteev, Peter",
		title          = "{Quantum deformations of the one-dimensional Hubbard model}",
		journal        = "J. Phys.",
		volume         = "A41",
		year           = "2008",
		pages          = "255204",
		doi            = "10.1088/1751-8113/41/25/255204",
		eprint         = "0802.0777",
		archivePrefix  = "arXiv",
		primaryClass   = "hep-th",
		reportNumber   = "AEI-2008-003, ITEP-TH-06-08",
		SLACcitation   = "
	}
	
	@article{Hoare:2011wr,
		author         = "Hoare, Ben and Hollowood, Timothy J. and Miramontes, J. Luis",
		title          = "{q-deformation of the $AdS_5 \times S^5$ superstring S-matrix and its relativistic limit}",
		journal        = "JHEP",
		volume         = "03",
		year           = "2012",
		pages          = "015",
		doi            = "10.1007/JHEP03(2012)015",
		eprint         = "1112.4485",
		archivePrefix  = "arXiv",
		primaryClass   = "hep-th",
		reportNumber   = "IMPERIAL-TP-11-BH-03",
		SLACcitation   = "
	}
	
	@article{Hoare:2014pna,
		author         = "Hoare, B. and Roiban, R. and Tseytlin, A. A.",
		title          = "{On deformations of $AdS_n \times S^n$ supercosets}",
		journal        = "JHEP",
		volume         = "06",
		year           = "2014",
		pages          = "002",
		doi            = "10.1007/JHEP06(2014)002",
		eprint         = "1403.5517",
		archivePrefix  = "arXiv",
		primaryClass   = "hep-th",
		reportNumber   = "IMPERIAL-TP-AT-2014-02, HU-EP-14-10",
		SLACcitation   = "
	}
	
	@article{Lunin:2014tsa,
		author         = "Lunin, O. and Roiban, R. and Tseytlin, A. A.",
		title          = "{Supergravity backgrounds for deformations of $AdS_n \times S^n$ supercoset string models}",
		journal        = "Nucl. Phys.",
		volume         = "B891",
		year           = "2015",
		pages          = "106-127",
		doi            = "10.1016/j.nuclphysb.2014.12.006",
		eprint         = "1411.1066",
		archivePrefix  = "arXiv",
		primaryClass   = "hep-th",
		reportNumber   = "IMPERIAL-TP-AT-2014-07",
		SLACcitation   = "
	}
	
	@article{Delduc:2013qra,
		author         = "Delduc, Francois and Magro, Marc and Vicedo, Benoit",
		title          = "{An integrable deformation of the $AdS_5 \times S^5$ superstring action}",
		journal        = "Phys. Rev. Lett.",
		volume         = "112",
		year           = "2014",
		number         = "5",
		pages          = "051601",
		doi            = "10.1103/PhysRevLett.112.051601",
		eprint         = "1309.5850",
		archivePrefix  = "arXiv",
		primaryClass   = "hep-th",
		SLACcitation   = "
	}
	
	@article{Delduc:2014kha,
		author         = "Delduc, Francois and Magro, Marc and Vicedo, Benoit",
		title          = "{Derivation of the action and symmetries of the $q$-deformed $AdS_5 \times S^5$ superstring}",
		journal        = "JHEP",
		volume         = "10",
		year           = "2014",
		pages          = "132",
		doi            = "10.1007/JHEP10(2014)132",
		eprint         = "1406.6286",
		archivePrefix  = "arXiv",
		primaryClass   = "hep-th",
		SLACcitation   = "
	}
	
	@article{Sorokin:2011rr,
		author         = "Sorokin, Dmitri and Tseytlin, Arkady and Wulff, Linus and Zarembo, Konstantin",
		title          = "{Superstrings in $AdS_2 \times S^2 \times T^6$}",
		journal        = "J. Phys.",
		volume         = "A44",
		year           = "2011",
		pages          = "275401",
		doi            = "10.1088/1751-8113/44/27/275401",
		eprint         = "1104.1793",
		archivePrefix  = "arXiv",
		primaryClass   = "hep-th",
		reportNumber   = "MIFPA-11-11, NORDITA-2011-30, IMPERIAL-TP-AT-2011-2",
		SLACcitation   = "
	}
	
	@article{Hoare:2016ibq,
		author         = "Hoare, Ben and van Tongeren, Stijn J.",
		title          = "{Non-split and split deformations of $AdS_5$}",
		journal        = "J. Phys.",
		volume         = "A49",
		year           = "2016",
		number         = "48",
		pages          = "484003",
		doi            = "10.1088/1751-8113/49/48/484003",
		eprint         = "1605.03552",
		archivePrefix  = "arXiv",
		primaryClass   = "hep-th",
		SLACcitation   = "
	}
	
	@article{Araujo:2018rbc,
		author         = "Araujo, Thiago and Colg\'{a}in, Eoin \'{O}. and Yavartanoo, Hossein",
		title          = "{Embedding the modified CYBE in Supergravity}",
		journal        = "Eur. Phys. J.",
		volume         = "C78",
		year           = "2018",
		number         = "10",
		pages          = "854",
		doi            = "10.1140/epjc/s10052-018-6335-6",
		eprint         = "1806.02602",
		archivePrefix  = "arXiv",
		primaryClass   = "hep-th",
		reportNumber   = "APCTP Pre2018-004, APCTP-PRE2018-004",
		SLACcitation   = "
	}
	
	@article{Arutyunov:2009ga,
		author         = "Arutyunov, Gleb and Frolov, Sergey",
		title          = "{Foundations of the $AdS_5 \times S^5$ Superstring. Part I}",
		journal        = "J. Phys.",
		volume         = "A42",
		year           = "2009",
		pages          = "254003",
		doi            = "10.1088/1751-8113/42/25/254003",
		eprint         = "0901.4937",
		archivePrefix  = "arXiv",
		primaryClass   = "hep-th",
		reportNumber   = "ITP-UU-09-05, SPIN-09-05, TCD-MATH-09-06, HMI-09-03",
		SLACcitation   = "
	}
	
	@article{Grigoriev:2007bu,
		author         = "Grigoriev, M. and Tseytlin, Arkady A.",
		title          = "{Pohlmeyer reduction of $AdS_5 \times S^5$ superstring sigma model}",
		journal        = "Nucl. Phys.",
		volume         = "B800",
		year           = "2008",
		pages          = "450-501",
		doi            = "10.1016/j.nuclphysb.2008.01.006",
		eprint         = "0711.0155",
		archivePrefix  = "arXiv",
		primaryClass   = "hep-th",
		reportNumber   = "IMPERIAL-TP-AT-2007-4",
		SLACcitation   = "
	}
	
	@article{Hoare:2015gda,
		author         = "Hoare, B. and Tseytlin, A. A.",
		title          = "{On integrable deformations of superstring sigma models related to $AdS_n \times S^n$ supercosets}",
		journal        = "Nucl. Phys.",
		volume         = "B897",
		year           = "2015",
		pages          = "448-478",
		doi            = "10.1016/j.nuclphysb.2015.06.001",
		eprint         = "1504.07213",
		archivePrefix  = "arXiv",
		primaryClass   = "hep-th",
		reportNumber   = "IMPERIAL-TP-AT-2015-02, HU-EP-15-21",
		SLACcitation   = "
	}
	
	@article{Klimcik:2002zj,
		author         = "Klim\v{c}\'{i}k, Ctirad",
		title          = "{Yang-Baxter $\sigma$-models and dS/AdS T-duality}",
		journal        = "JHEP",
		volume         = "12",
		year           = "2002",
		pages          = "051",
		doi            = "10.1088/1126-6708/2002/12/051",
		eprint         = "hep-th/0210095",
		archivePrefix  = "arXiv",
		primaryClass   = "hep-th",
		reportNumber   = "IML-02-XY",
		SLACcitation   = "
	}
	
	@article{Delduc:2013fga,
		author         = "Delduc, Francois and Magro, Marc and Vicedo, Benoit",
		title          = "{On classical $q$-deformations of integrable $\sigma$-models}",
		journal        = "JHEP",
		volume         = "11",
		year           = "2013",
		pages          = "192",
		doi            = "10.1007/JHEP11(2013)192",
		eprint         = "1308.3581",
		archivePrefix  = "arXiv",
		primaryClass   = "hep-th",
		SLACcitation   = "
	}
	
	@article{Bena:2003wd,
		author         = "Bena, Iosif and Polchinski, Joseph and Roiban, Radu",
		title          = "{Hidden symmetries of the $AdS_5 \times S^5$ superstring}",
		journal        = "Phys. Rev.",
		volume         = "D69",
		year           = "2004",
		pages          = "046002",
		doi            = "10.1103/PhysRevD.69.046002",
		eprint         = "hep-th/0305116",
		archivePrefix  = "arXiv",
		primaryClass   = "hep-th",
		reportNumber   = "NSF-KITP-03-34, UCLA-03-TEP-14",
		SLACcitation   = "
	}
	
	@article{Magro:2008dv,
		author         = "Magro, Marc",
		title          = "{The Classical Exchange Algebra of $AdS_5 \times S^5$}",
		journal        = "JHEP",
		volume         = "01",
		year           = "2009",
		pages          = "021",
		doi            = "10.1088/1126-6708/2009/01/021",
		eprint         = "0810.4136",
		archivePrefix  = "arXiv",
		primaryClass   = "hep-th",
		reportNumber   = "AEI-2008-085",
		SLACcitation   = "
	}
	
	@article{Vicedo:2010qd,
		author         = "Vicedo, Benoit",
		title          = "{The classical R-matrix of AdS/CFT and its Lie dialgebra structure}",
		journal        = "Lett. Math. Phys.",
		volume         = "95",
		year           = "2011",
		pages          = "249-274",
		doi            = "10.1007/s11005-010-0446-9",
		eprint         = "1003.1192",
		archivePrefix  = "arXiv",
		primaryClass   = "hep-th",
		reportNumber   = "IPHT-T10-026",
		SLACcitation   = "
	}
	
	@article{Severa:2018pag,
		author         = "\v{S}evera, Pavol and Valach, Fridrich",
		title          = "{Courant algebroids, Poisson-Lie T-duality, and type II supergravities}",
		year           = "2018",
		eprint         = "1810.07763",
		archivePrefix  = "arXiv",
		primaryClass   = "math.DG",
		SLACcitation   = "
	}
	
	@article{Demulder:2018lmj,
		author         = "Demulder, Saskia and Hassler, Falk and Thompson, Daniel C.",
		title          = "{Doubled aspects of generalised dualities and integrable deformations}",
		year           = "2018",
		eprint         = "1810.11446",
		archivePrefix  = "arXiv",
		primaryClass   = "hep-th",
		SLACcitation   = "
	}
	
	@article{Araujo:2017enj,
		author         = "Araujo, T. and Colg\'{a}in, E. \'{O} and Sakamoto, J. and Sheikh-Jabbari, M. M. and Yoshida, K.",
		title          = "{$I$ in generalized supergravity}",
		journal        = "Eur. Phys. J.",
		volume         = "C77",
		year           = "2017",
		number         = "11",
		pages          = "739",
		doi            = "10.1140/epjc/s10052-017-5316-5",
		eprint         = "1708.03163",
		archivePrefix  = "arXiv",
		primaryClass   = "hep-th",
		reportNumber   = "APCTP-PRE2017---015, KUNS-2696, IPM-P-2017-024,IPM-P-2017-028",
		SLACcitation   = "
	}
	
	@article{Hoare:2014oua,
		author         = "Hoare, Ben",
		title          = "{Towards a two-parameter q-deformation of $AdS_3 \times S^3 \times M^4$ superstrings}",
		journal        = "Nucl. Phys.",
		volume         = "B891",
		year           = "2015",
		pages          = "259-295",
		doi            = "10.1016/j.nuclphysb.2014.12.012",
		eprint         = "1411.1266",
		archivePrefix  = "arXiv",
		primaryClass   = "hep-th",
		reportNumber   = "HU-EP-14-44",
		SLACcitation   = "
	}
	
	@article{Delduc:2018xug,
		author         = "Delduc, Francois and Hoare, Ben and Kameyama, Takashi and Lacroix, Sylvain and Magro, Marc",
		title          = "{Three-parameter integrable deformation of $\Integer_4$ permutation supercosets}",
		year           = "2018",
		eprint         = "1811.00453",
		archivePrefix  = "arXiv",
		primaryClass   = "hep-th",
		reportNumber   = "ZMP-HH/18-22",
		SLACcitation   = "
	}
	
	@article{Cagnazzo:2012se,
		author         = "Cagnazzo, A. and Zarembo, K.",
		title          = "{B-field in $AdS_3/CFT_2$ Correspondence and Integrability}",
		journal        = "JHEP",
		volume         = "11",
		year           = "2012",
		pages          = "133",
		doi            = "10.1007/JHEP11(2012)133",
		note           = "[Erratum: \doiref{10.1007/JHEP04(2013)003}{\textsf{JHEP 1304, 003 (2013)}}]",
		eprint         = "1209.4049",
		archivePrefix  = "arXiv",
		primaryClass   = "hep-th",
		reportNumber   = "NORDITA-2012-67, UUITP-24-12",
		SLACcitation   = "
	}
	
	@article{Hoare:2018ebg,
		author         = "Hoare, Ben and Seibold, Fiona K.",
		title          = "{Poisson-Lie duals of the $\eta$-deformed $AdS_2 \times S^2 \times T^6$ superstring}",
		journal        = "JHEP",
		volume         = "08",
		year           = "2018",
		pages          = "107",
		doi            = "10.1007/JHEP08(2018)107",
		eprint         = "1807.04608",
		archivePrefix  = "arXiv",
		primaryClass   = "hep-th",
		SLACcitation   = "
	}
	
	@article{Alvarez:1994np,
		author         = "\'{A}lvarez, Enrique and \'{A}lvarez-Gaum\'{e}, Luis and Lozano, Yolanda",
		title          = "{On non-abelian duality}",
		journal        = "Nucl. Phys.",
		volume         = "B424",
		year           = "1994",
		pages          = "155-183",
		doi            = "10.1016/0550-3213(94)90093-0",
		eprint         = "hep-th/9403155",
		archivePrefix  = "arXiv",
		primaryClass   = "hep-th",
		reportNumber   = "CERN-TH-7204-94",
		SLACcitation   = "
	}
	
	@article{Elitzur:1994ri,
		author         = "Elitzur, S. and Giveon, A. and Rabinovici, E. and Schwimmer, A. and Veneziano, G.",
		title          = "{Remarks on non-abelian duality}",
		journal        = "Nucl. Phys.",
		volume         = "B435",
		year           = "1995",
		pages          = "147-171",
		doi            = "10.1016/0550-3213(94)00426-F",
		eprint         = "hep-th/9409011",
		archivePrefix  = "arXiv",
		primaryClass   = "hep-th",
		reportNumber   = "CERN-TH-7414-94, RI-9-94, WIS-7-94",
		SLACcitation   = "
	}
	
	@article{Tyurin:1995bu,
		author         = "Tyurin, Eugene and von Unge, Rikard",
		title          = "{Poisson-Lie T-duality: the path-integral derivation}",
		journal        = "Phys. Lett.",
		volume         = "B382",
		year           = "1996",
		pages          = "233-240",
		doi            = "10.1016/0370-2693(96)00680-6",
		eprint         = "hep-th/9512025",
		archivePrefix  = "arXiv",
		primaryClass   = "hep-th",
		reportNumber   = "ITP-SB-95-50, USITP-95-11",
		SLACcitation   = "
	}
	
	@article{Bossard:2001au,
		author         = "Bossard, A. and Mohammedi, N.",
		title          = "{Poisson-Lie duality in the string effective action}",
		journal        = "Nucl. Phys.",
		volume         = "B619",
		year           = "2001",
		pages          = "128-154",
		doi            = "10.1016/S0550-3213(01)00541-7",
		eprint         = "hep-th/0106211",
		archivePrefix  = "arXiv",
		primaryClass   = "hep-th",
		SLACcitation   = "
	}
	
	@article{Zarembo:2010sg,
		author         = "Zarembo, K.",
		title          = "{Strings on semisymmetric superspaces}",
		journal        = "JHEP",
		volume         = "05",
		year           = "2010",
		pages          = "002",
		doi            = "10.1007/JHEP05(2010)002",
		eprint         = "1003.0465",
		archivePrefix  = "arXiv",
		primaryClass   = "hep-th",
		reportNumber   = "ITEP-TH-12-10, LPTENS-10-12, UUITP-05-10",
		SLACcitation   = "
	}
	
	@article{Fateev:1992tk,
		author         = "Fateev, V. A. and Onofri, E. and Zamolodchikov, Alexei B.",
		title          = "{The sausage model (integrable deformations of O(3) sigma model)}",
		journal        = "Nucl. Phys.",
		volume         = "B406",
		year           = "1993",
		pages          = "521-565",
		doi            = "10.1016/0550-3213(93)90001-6",
		reportNumber   = "PAR-LPTHE-92-46, LPTHE-92-46",
		SLACcitation   = "
	}
	
	@article{Klimcik:2008eq,
		author         = "Klim\v{c}\'{i}k, Ctirad",
		title          = "{On integrability of the Yang-Baxter $\sigma$-model}",
		journal        = "J. Math. Phys.",
		volume         = "50",
		year           = "2009",
		pages          = "043508",
		doi            = "10.1063/1.3116242",
		eprint         = "0802.3518",
		archivePrefix  = "arXiv",
		primaryClass   = "hep-th",
		SLACcitation   = "
	}
	
	@article{Klimcik:2014bta,
		author         = "Klim\v{c}\'{i}k, Ctirad",
		title          = "{Integrability of the bi-Yang-Baxter sigma-model}",
		journal        = "Lett. Math. Phys.",
		volume         = "104",
		year           = "2014",
		pages          = "1095-1106",
		doi            = "10.1007/s11005-014-0709-y",
		eprint         = "1402.2105",
		archivePrefix  = "arXiv",
		primaryClass   = "math-ph",
		SLACcitation   = "
	}
	
	@article{Drinfeld:1985rx,
		author         = "Drinfel'd, V. G.",
		title          = "{Hopf algebras and the quantum Yang-Baxter equation}",
		journal        = "Sov. Math. Dokl.",
		volume         = "32",
		year           = "1985",
		pages          = "254",
		note           = "[\textsf{Dokl. Akad. Nauk Ser. Fiz. 283, 1060 (1985)}]",
		SLACcitation   = "
	}
	
	@article{Jimbo:1985zk,
		author         = "Jimbo, Michio",
		title          = "{A q-difference analog of U(g) and the Yang-Baxter equation}",
		journal        = "Lett. Math. Phys.",
		volume         = "10",
		year           = "1985",
		pages          = "63",
		doi            = "10.1007/BF00704588",
		SLACcitation   = "
	}
	
	@article{Belavin:1984,
		author         = "Belavin, A. A. and Drinfel'd, V. G.",
		title          = "{Triangle equations and simple Lie algebras}",
		journal        = "Sov. Sci. Rev.",
		volume         = "C4",
		year           = "1984",
		pages          = "93",
	}
	
	@article{SemenovTianShansky:1983ik,
		author         = "Semenov-Tian-Shansky, M. A.",
		title          = "{What is a classical r-matrix?}",
		journal        = "Funct. Anal. Appl.",
		volume         = "17",
		year           = "1983",
		pages          = "259-272",
		doi            = "10.1007/BF01076717",
		note           = "[\textsf{Funkt. Anal. Pril. 17, 17 (1983)}]",
		SLACcitation   = "
	}
	
	@article{Klimcik:1995ux,
		author         = "Klim\v{c}\'{i}k, C. and \v{S}evera, P.",
		title          = "{Dual non-abelian duality and the Drinfel'd double}",
		journal        = "Phys. Lett.",
		volume         = "B351",
		year           = "1995",
		pages          = "455-462",
		doi            = "10.1016/0370-2693(95)00451-P",
		eprint         = "hep-th/9502122",
		archivePrefix  = "arXiv",
		primaryClass   = "hep-th",
		reportNumber   = "CERN-TH-95-39, CERN-TH-95-039",
		SLACcitation   = "
	}
	
	@article{Klimcik:1995jn,
		author         = "Klim\v{c}\'{i}k, C.",
		title          = "{Poisson-Lie T-duality}",
		journal        = "Nucl. Phys. Proc. Suppl.",
		volume         = "46",
		year           = "1996",
		pages          = "116-121",
		doi            = "10.1016/0920-5632(96)00013-8",
		eprint         = "hep-th/9509095",
		archivePrefix  = "arXiv",
		primaryClass   = "hep-th",
		reportNumber   = "CERN-TH-95-248",
		SLACcitation   = "
	}
	
	@article{Klimcik:1995dy,
		author         = "Klim\v{c}\'{i}k, C. and \v{S}evera, P.",
		title          = "{Poisson-Lie T-duality and loop groups of Drinfel'd doubles}",
		journal        = "Phys. Lett.",
		volume         = "B372",
		year           = "1996",
		pages          = "65-71",
		doi            = "10.1016/0370-2693(96)00025-1",
		eprint         = "hep-th/9512040",
		archivePrefix  = "arXiv",
		primaryClass   = "hep-th",
		reportNumber   = "CERN-TH-95-330",
		SLACcitation   = "
	}
	
	@article{Klimcik:1996nq,
		author         = "Klim\v{c}\'{i}k, C. and \v{S}evera, P.",
		title          = "{Non-abelian momentum-winding exchange}",
		journal        = "Phys. Lett.",
		volume         = "B383",
		year           = "1996",
		pages          = "281-286",
		doi            = "10.1016/0370-2693(96)00755-1",
		eprint         = "hep-th/9605212",
		archivePrefix  = "arXiv",
		primaryClass   = "hep-th",
		reportNumber   = "CERN-TH-96-142",
		SLACcitation   = "
	}
	
	@article{Delduc:2016ihq,
		author         = "Delduc, Francois and Lacroix, Sylvain and Magro, Marc and Vicedo, Benoit",
		title          = "{On $q$-deformed symmetries as Poisson-Lie symmetries and application to Yang-Baxter type models}",
		journal        = "J. Phys.",
		volume         = "A49",
		year           = "2016",
		number         = "41",
		pages          = "415402",
		doi            = "10.1088/1751-8113/49/41/415402",
		eprint         = "1606.01712",
		archivePrefix  = "arXiv",
		primaryClass   = "hep-th",
		SLACcitation   = "
	}
	
	@article{Green:1983wt,
		author         = "Green, Michael B. and Schwarz, John H.",
		title          = "{Covariant Description of Superstrings}",
		journal        = "Phys. Lett.",
		volume         = "B136",
		year           = "1984",
		pages          = "367",
		doi            = "10.1016/0370-2693(84)92021-5",
		reportNumber   = "QMC-83-7",
		SLACcitation   = "
	}
	
	@article{Grisaru:1985fv,
		author         = "Grisaru, Marcus T. and Howe, Paul S. and Mezincescu, L. and Nilsson, B. and Townsend, P. K.",
		title          = "{N=2 Superstrings in a Supergravity Background}",
		journal        = "Phys. Lett.",
		volume         = "B162",
		year           = "1985",
		pages          = "116",
		doi            = "10.1016/0370-2693(85)91071-8",
		reportNumber   = "Print-85-0603 (CAMBRIDGE)",
		SLACcitation   = "
	}
	
	@article{Tseytlin:1996hs,
		author         = "Tseytlin, Arkady A.",
		title          = "{On dilaton dependence of type II superstring action}",
		journal        = "Class. Quant. Grav.",
		volume         = "13",
		year           = "1996",
		pages          = "L81",
		doi            = "10.1088/0264-9381/13/6/003",
		eprint         = "hep-th/9601109",
		archivePrefix  = "arXiv",
		primaryClass   = "hep-th",
		reportNumber   = "IMPERIAL-TP-95-96-19",
		SLACcitation   = "
	}
	
	@article{Cvetic:1999zs,
		author         = "Cveti\v{c}, Mirjam and Lu, Hong and Pope, C. N. and Stelle, K. S.",
		title          = "{T-duality in the Green-Schwarz formalism, and the massless/massive IIA duality map}",
		journal        = "Nucl. Phys.",
		volume         = "B573",
		year           = "2000",
		pages          = "149",
		doi            = "10.1016/S0550-3213(99)00740-3",
		eprint         = "hep-th/9907202",
		archivePrefix  = "arXiv",
		primaryClass   = "hep-th",
		reportNumber   = "UPR-0852-T, CTP-TAMU-31-99, SISSA-88-99-EP,
		IMPERIAL-TP-98-99-63, NSF-ITP-99-086",
		SLACcitation   = "
	}
	
	@article{Wulff:2013kga,
		author         = "Wulff, Linus",
		title          = "{The type II superstring to order $\theta^4$}",
		journal        = "JHEP",
		volume         = "07",
		year           = "2013",
		pages          = "123",
		doi            = "10.1007/JHEP07(2013)123",
		eprint         = "1304.6422",
		archivePrefix  = "arXiv",
		primaryClass   = "hep-th",
		reportNumber   = "MIFPA-13-13",
		SLACcitation   = "
	}
	
	@article{Borowiec:2015wua,
		author         = "Borowiec, Andrzej and Kyono, Hideki and Lukierski, Jerzy and Sakamoto, Jun-ichi and Yoshida, Kentaroh",
		title          = "{Yang-Baxter sigma models and Lax pairs arising from $\kappa$-Poincar\'{e} $r$-matrices}",
		journal        = "JHEP",
		volume         = "04",
		year           = "2016",
		pages          = "079",
		doi            = "10.1007/JHEP04(2016)079",
		eprint         = "1510.03083",
		archivePrefix  = "arXiv",
		primaryClass   = "hep-th",
		reportNumber   = "KUNS-2578",
		SLACcitation   = "
	}
	
	@article{Wulff:2014kja,
		author         = "Wulff, Linus",
		title          = "{Superisometries and integrability of superstrings}",
		journal        = "JHEP",
		volume         = "05",
		year           = "2014",
		pages          = "115",
		doi            = "10.1007/JHEP05(2014)115",
		eprint         = "1402.3122",
		archivePrefix  = "arXiv",
		primaryClass   = "hep-th",
		reportNumber   = "IMPERIAL-TP-LW-2014-01",
		SLACcitation   = "
	}
	
	@article{Borsato:2018idb,
		author         = "Borsato, Riccardo and Wulff, Linus",
		title          = "{Non-abelian T-duality and Yang-Baxter deformations of Green-Schwarz strings}",
		journal        = "JHEP",
		volume         = "08",
		year           = "2018",
		pages          = "027",
		doi            = "10.1007/JHEP08(2018)027",
		eprint         = "1806.04083",
		archivePrefix  = "arXiv",
		primaryClass   = "hep-th",
		reportNumber   = "NORDITA 2018-041, NORDITA-2018-041",
		SLACcitation   = "
	}
	
	@article{Roychowdhury:2018qsz,
		author         = "Roychowdhury, Dibakar",
		title          = "{On pp wave limit for $\eta$ deformed superstrings}",
		journal        = "JHEP",
		volume         = "05",
		year           = "2018",
		pages          = "018",
		doi            = "10.1007/JHEP05(2018)018",
		eprint         = "1801.07680",
		archivePrefix  = "arXiv",
		primaryClass   = "hep-th",
		SLACcitation   = "
	}
	
	@article{Berenstein:2002jq,
		author         = "Berenstein, David Eliecer and Maldacena, Juan Martin and Nastase, Horatiu Stefan",
		title          = "{Strings in flat space and pp waves from $\mathcal{N} = 4$ Super Yang Mills}",
		journal        = "JHEP",
		volume         = "04",
		year           = "2002",
		pages          = "013",
		doi            = "10.1088/1126-6708/2002/04/013",
		eprint         = "hep-th/0202021",
		archivePrefix  = "arXiv",
		primaryClass   = "hep-th",
		SLACcitation   = "
	}
	
	@article{Blau:2002dy,
		author         = "Blau, Matthias and Figueroa-O'Farrill, Jose M. and Hull, Christopher and Papadopoulos, George",
		title          = "{Penrose limits and maximal supersymmetry}",
		journal        = "Class. Quant. Grav.",
		volume         = "19",
		year           = "2002",
		pages          = "L87-L95",
		doi            = "10.1088/0264-9381/19/10/101",
		eprint         = "hep-th/0201081",
		archivePrefix  = "arXiv",
		primaryClass   = "hep-th",
		reportNumber   = "EMPG-02-01, QMUL-PH-02-01",
		SLACcitation   = "
	}
	
	@article{Sakamoto:2017wor,
		author         = "Sakamoto, Jun-ichi and Sakatani, Yuho and Yoshida, Kentaroh",
		title          = "{Weyl invariance for generalized supergravity backgrounds from the doubled formalism}",
		journal        = "PTEP",
		volume         = "2017",
		year           = "2017",
		number         = "5",
		pages          = "053B07",
		doi            = "10.1093/ptep/ptx067",
		eprint         = "1703.09213",
		archivePrefix  = "arXiv",
		primaryClass   = "hep-th",
		reportNumber   = "KUNS-2668",
		SLACcitation   = "
	}
	
	@article{Fernandez-Melgarejo:2018wpg,
		author         = "Fern\'{a}ndez-Melgarejo, Jose J. and Sakamoto, Jun-ichi and Sakatani, Yuho and Yoshida, Kentaroh",
		title          = "{Comments on Weyl invariance of string theories in generalized supergravity backgrounds}",
		year           = "2018",
		eprint         = "1811.10600",
		archivePrefix  = "arXiv",
		primaryClass   = "hep-th",
		SLACcitation   = "
	}

@article{Maldacena:2000hw,
	author         = "Maldacena, Juan Martin and Ooguri, Hirosi",
	title          = "{Strings in AdS(3) and SL(2,R) WZW model 1.: The
		Spectrum}",
	journal        = "J. Math. Phys.",
	volume         = "42",
	year           = "2001",
	pages          = "2929-2960",
	doi            = "10.1063/1.1377273",
	eprint         = "hep-th/0001053",
	archivePrefix  = "arXiv",
	primaryClass   = "hep-th",
	reportNumber   = "CALT-68-2245, CITUSC-99-010, HUTP-99-A027, LBNL-44375,
	UCB-PTH-99-48, LBL-44375",
	SLACcitation   = "
}
@article{Maldacena:2000kv,
	author         = "Maldacena, Juan Martin and Ooguri, Hirosi and Son, John",
	title          = "{Strings in AdS(3) and the SL(2,R) WZW model. Part 2.
		Euclidean black hole}",
	journal        = "J. Math. Phys.",
	volume         = "42",
	year           = "2001",
	pages          = "2961-2977",
	doi            = "10.1063/1.1377039",
	eprint         = "hep-th/0005183",
	archivePrefix  = "arXiv",
	primaryClass   = "hep-th",
	reportNumber   = "CALT-68-2266, CITUSC-00-021, HUTP-00-A009, UCB-PTH-00-10",
	SLACcitation   = "
}
@article{Maldacena:2001km,
	author         = "Maldacena, Juan Martin and Ooguri, Hirosi",
	title          = "{Strings in AdS(3) and the SL(2,R) WZW model. Part 3.
		Correlation functions}",
	journal        = "Phys. Rev.",
	volume         = "D65",
	year           = "2002",
	pages          = "106006",
	doi            = "10.1103/PhysRevD.65.106006",
	eprint         = "hep-th/0111180",
	archivePrefix  = "arXiv",
	primaryClass   = "hep-th",
	reportNumber   = "CALT-68-2360, CITUSC-01-042",
	SLACcitation   = "
}

@article{Muck:2019pwj,
	author         = "Mück, Wolfgang",
	title          = "{Generalized Supergravity Equations and Generalized
		Fradkin-Tseytlin Counterterm}",
	journal        = "JHEP",
	volume         = "05",
	year           = "2019",
	pages          = "063",
	doi            = "10.1007/JHEP05(2019)063",
	eprint         = "1904.06126",
	archivePrefix  = "arXiv",
	primaryClass   = "hep-th",
	SLACcitation   = "
}

@article{Hoare:2009fs,
	author         = "Hoare, B. and Tseytlin, A. A.",
	title          = "{Tree-level S-matrix of Pohlmeyer reduced form of $AdS_5 \times  S^5$ superstring theory}",
	journal        = "JHEP",
	volume         = "02",
	year           = "2010",
	pages          = "094",
	doi            = "10.1007/JHEP02(2010)094",
	eprint         = "0912.2958",
	archivePrefix  = "arXiv",
	primaryClass   = "hep-th",
	reportNumber   = "IMPERIAL-TP-AT-2009-06",
	SLACcitation   = "
}
@article{Hoare:2011fj,
	author         = "Hoare, B. and Tseytlin, A. A.",
	title          = "{Towards the quantum S-matrix of the Pohlmeyer reduced
		version of $AdS_5 \times  S^5$ superstring theory}",
	journal        = "Nucl. Phys.",
	volume         = "B851",
	year           = "2011",
	pages          = "161-237",
	doi            = "10.1016/j.nuclphysb.2011.05.016",
	eprint         = "1104.2423",
	archivePrefix  = "arXiv",
	primaryClass   = "hep-th",
	reportNumber   = "IMPERIAL-TP-BH-2011-01",
	SLACcitation   = "
}@article{Beisert:2010kk,
author         = "Beisert, Niklas",
title          = "{The Classical Trigonometric r-Matrix for the
	Quantum-Deformed Hubbard Chain}",
journal        = "J. Phys.",
volume         = "A44",
year           = "2011",
pages          = "265202",
doi            = "10.1088/1751-8113/44/26/265202",
eprint         = "1002.1097",
archivePrefix  = "arXiv",
primaryClass   = "math-ph",
reportNumber   = "AEI-2010-016",
SLACcitation   = "
}
@article{Delduc:2018xug,
	author         = "Delduc, F. and Hoare, B. and Kameyama, T. and Lacroix, S.
	and Magro, M.",
	title          = "{Three-parameter integrable deformation of $\mathbb{Z}_4$
		permutation supercosets}",
	journal        = "JHEP",
	volume         = "01",
	year           = "2019",
	pages          = "109",
	doi            = "10.1007/JHEP01(2019)109",
	eprint         = "1811.00453",
	archivePrefix  = "arXiv",
	primaryClass   = "hep-th",
	reportNumber   = "ZMP-HH/18-22",
	SLACcitation   = "
}
@article{Giribet:2018ada,
	author         = "Giribet, G. and Hull, C. and Kleban, M. and Porrati, M.
	and Rabinovici, E.",
	title          = "{Superstrings on AdS$_{3}$ at $k= 1$}",
	journal        = "JHEP",
	volume         = "08",
	year           = "2018",
	pages          = "204",
	doi            = "10.1007/JHEP08(2018)204",
	eprint         = "1803.04420",
	archivePrefix  = "arXiv",
	primaryClass   = "hep-th",
	reportNumber   = "Imperial-TP-2018-CH-01, IMPERIAL-TP-2018-CH-01",
	SLACcitation   = "
}
@article{Gaberdiel:2018rqv,
	author         = "Gaberdiel, Matthias R. and Gopakumar, Rajesh",
	title          = "{Tensionless string spectra on AdS$_{3}$}",
	journal        = "JHEP",
	volume         = "05",
	year           = "2018",
	pages          = "085",
	doi            = "10.1007/JHEP05(2018)085",
	eprint         = "1803.04423",
	archivePrefix  = "arXiv",
	primaryClass   = "hep-th",
	SLACcitation   = "
}
@article{Eberhardt:2018ouy,
	author         = "Eberhardt, Lorenz and Gaberdiel, Matthias R. and
	Gopakumar, Rajesh",
	title          = "{The Worldsheet Dual of the Symmetric Product CFT}",
	journal        = "JHEP",
	volume         = "04",
	year           = "2019",
	pages          = "103",
	doi            = "10.1007/JHEP04(2019)103",
	eprint         = "1812.01007",
	archivePrefix  = "arXiv",
	primaryClass   = "hep-th",
	SLACcitation   = "
}
@article{Eberhardt:2019qcl,
	author         = "Eberhardt, Lorenz and Gaberdiel, Matthias R.",
	title          = "{String theory on $\boldsymbol{\text{AdS}_{\mathbf{3}}}$
		and the symmetric orbifold of Liouville theory}",
	year           = "2019",
	eprint         = "1903.00421",
	archivePrefix  = "arXiv",
	primaryClass   = "hep-th",
	SLACcitation   = "
}

@article{Maldacena:1997re,
	author         = "Maldacena, Juan Martin",
	title          = "{The Large N limit of superconformal field theories and
		supergravity}",
	journal        = "Int. J. Theor. Phys.",
	volume         = "38",
	year           = "1999",
	pages          = "1113-1133",
	doi            = "10.1023/A:1026654312961, 10.4310/ATMP.1998.v2.n2.a1",
	note           = "[Adv. Theor. Math. Phys.2,231(1998)]",
	eprint         = "hep-th/9711200",
	archivePrefix  = "arXiv",
	primaryClass   = "hep-th",
	reportNumber   = "HUTP-97-A097, HUTP-98-A097",
	SLACcitation   = "
}
@article{Aharony:1999ti,
	author         = "Aharony, Ofer and Gubser, Steven S. and Maldacena, Juan
	Martin and Ooguri, Hirosi and Oz, Yaron",
	title          = "{Large N field theories, string theory and gravity}",
	journal        = "Phys. Rept.",
	volume         = "323",
	year           = "2000",
	pages          = "183-386",
	doi            = "10.1016/S0370-1573(99)00083-6",
	eprint         = "hep-th/9905111",
	archivePrefix  = "arXiv",
	primaryClass   = "hep-th",
	reportNumber   = "CERN-TH-99-122, HUTP-99-A027, LBNL-43113, RU-99-18,
	UCB-PTH-99-16, LBL-43113",
	SLACcitation   = "
}
@article{Banados:1992wn,
	author         = "Banados, Maximo and Teitelboim, Claudio and Zanelli,
	Jorge",
	title          = "{The Black hole in three-dimensional space-time}",
	journal        = "Phys. Rev. Lett.",
	volume         = "69",
	year           = "1992",
	pages          = "1849-1851",
	doi            = "10.1103/PhysRevLett.69.1849",
	eprint         = "hep-th/9204099",
	archivePrefix  = "arXiv",
	primaryClass   = "hep-th",
	reportNumber   = "PRINT-92-0151 (CHILE), IASSNS-HEP-92-29",
	SLACcitation   = "
}
@article{Bykov:2016pfu,
	author         = "Bykov, Dmitri",
	title          = "{Complex structure-induced deformations of sigma-models}",
	journal        = "JHEP",
	volume         = "03",
	year           = "2017",
	pages          = "130",
	doi            = "10.1007/JHEP03(2017)130",
	eprint         = "1611.07116",
	archivePrefix  = "arXiv",
	primaryClass   = "hep-th",
	SLACcitation   = "
}
@article{Arutyunov:2007tc,
	author         = "Arutyunov, Gleb and Frolov, Sergey",
	title          = "{On String S-matrix, Bound States and TBA}",
	journal        = "JHEP",
	volume         = "12",
	year           = "2007",
	pages          = "024",
	doi            = "10.1088/1126-6708/2007/12/024",
	eprint         = "0710.1568",
	archivePrefix  = "arXiv",
	primaryClass   = "hep-th",
	reportNumber   = "ITP-UU-07-50, SPIN-07-37, TCDMATH-07-15",
	SLACcitation   = "
}
@article{Ambjorn:2005wa,
	author         = "Ambjorn, Jan and Janik, Romuald A. and Kristjansen,
	Charlotte",
	title          = "{Wrapping interactions and a new source of corrections to
		the spin-chain/string duality}",
	journal        = "Nucl. Phys.",
	volume         = "B736",
	year           = "2006",
	pages          = "288-301",
	doi            = "10.1016/j.nuclphysb.2005.12.007",
	eprint         = "hep-th/0510171",
	archivePrefix  = "arXiv",
	primaryClass   = "hep-th",
	reportNumber   = "NORDITA-2005-67",
	SLACcitation   = "
}

@article{Klimcik:2008eq,
	author         = "Klimcik, Ctirad",
	title          = "{On integrability of the Yang-Baxter sigma-model}",
	journal        = "J. Math. Phys.",
	volume         = "50",
	year           = "2009",
	pages          = "043508",
	doi            = "10.1063/1.3116242",
	eprint         = "0802.3518",
	archivePrefix  = "arXiv",
	primaryClass   = "hep-th",
	SLACcitation   = "
}
@article{Klimcik:2014bta,
	author         = "Klimcik, Ctirad",
	title          = "{Integrability of the bi-Yang-Baxter sigma-model}",
	journal        = "Lett. Math. Phys.",
	volume         = "104",
	year           = "2014",
	pages          = "1095-1106",
	doi            = "10.1007/s11005-014-0709-y",
	eprint         = "1402.2105",
	archivePrefix  = "arXiv",
	primaryClass   = "math-ph",
	SLACcitation   = "
}
@article{Berenstein:2002jq,
	author         = "Berenstein, David Eliecer and Maldacena, Juan Martin and
	Nastase, Horatiu Stefan",
	title          = "{Strings in flat space and pp waves from N=4
		superYang-Mills}",
	journal        = "JHEP",
	volume         = "04",
	year           = "2002",
	pages          = "013",
	doi            = "10.1088/1126-6708/2002/04/013",
	eprint         = "hep-th/0202021",
	archivePrefix  = "arXiv",
	primaryClass   = "hep-th",
	SLACcitation   = "
}
@article{David:2002wn,
	author         = "David, Justin R. and Mandal, Gautam and Wadia, Spenta R.",
	title          = "{Microscopic formulation of black holes in string
		theory}",
	journal        = "Phys. Rept.",
	volume         = "369",
	year           = "2002",
	pages          = "549-686",
	doi            = "10.1016/S0370-1573(02)00271-5",
	eprint         = "hep-th/0203048",
	archivePrefix  = "arXiv",
	primaryClass   = "hep-th",
	reportNumber   = "TIFR-TH-02-07",
	SLACcitation   = "
}
@article{Sfondrini:2014via,
	author         = "Sfondrini, Alessandro",
	title          = "{Towards integrability for ${\rm Ad}{{{\rm S}}_{{\bf
					3}}}/{\rm CF}{{{\rm T}}_{{\bf 2}}}$}",
	journal        = "J. Phys.",
	volume         = "A48",
	year           = "2015",
	number         = "2",
	pages          = "023001",
	doi            = "10.1088/1751-8113/48/2/023001",
	eprint         = "1406.2971",
	archivePrefix  = "arXiv",
	primaryClass   = "hep-th",
	reportNumber   = "HU-MATHEMATIK-2014-14, HU-EP-14-24",
	SLACcitation   = "
}
@article{Babichenko:2009dk,
	author         = "Babichenko, A. and Stefanski, Jr., B. and Zarembo, K.",
	title          = "{Integrability and the AdS(3)/CFT(2) correspondence}",
	journal        = "JHEP",
	volume         = "03",
	year           = "2010",
	pages          = "058",
	doi            = "10.1007/JHEP03(2010)058",
	eprint         = "0912.1723",
	archivePrefix  = "arXiv",
	primaryClass   = "hep-th",
	reportNumber   = "ITEP-TH-59-09, LPTENS-09-36, UUITP-25-09",
	SLACcitation   = "
}
@article{Borsato:2012ud,
	author         = "Borsato, Riccardo and Ohlsson Sax, Olof and Sfondrini,
	Alessandro",
	title          = "{A dynamic $su(1|1)^2$ S-matrix for AdS3/CFT2}",
	journal        = "JHEP",
	volume         = "04",
	year           = "2013",
	pages          = "113",
	doi            = "10.1007/JHEP04(2013)113",
	eprint         = "1211.5119",
	archivePrefix  = "arXiv",
	primaryClass   = "hep-th",
	reportNumber   = "ITP-UU-12-46, SPIN-12-43",
	SLACcitation   = "
}
@article{Hoare:2013lja,
	author         = "Hoare, B. and Stepanchuk, A. and Tseytlin, A. A.",
	title          = "{Giant magnon solution and dispersion relation in string
		theory in $AdS_3 \times S^3 \times T^4$ with mixed flux}",
	journal        = "Nucl. Phys.",
	volume         = "B879",
	year           = "2014",
	pages          = "318-347",
	doi            = "10.1016/j.nuclphysb.2013.12.011",
	eprint         = "1311.1794",
	archivePrefix  = "arXiv",
	primaryClass   = "hep-th",
	reportNumber   = "IMPERIAL-TP-AS-2013-01, HU-EP-13-56",
	SLACcitation   = "
}
@article{Hernandez:2014eta,
	author         = "Hernández, Rafael and Nieto, Juan Miguel",
	title          = "{Spinning strings in $AdS_3 \times S^3$ with NS-NS
		flux}",
	journal        = "Nucl. Phys.",
	volume         = "B888",
	year           = "2014",
	pages          = "236-247",
	doi            = "10.1016/j.nuclphysb.2015.04.011,
	10.1016/j.nuclphysb.2014.10.001",
	note           = "[Erratum: Nucl. Phys.B895,303(2015)]",
	eprint         = "1407.7475",
	archivePrefix  = "arXiv",
	primaryClass   = "hep-th",
	SLACcitation   = "
}
@article{Baggio:2017kza,
	author         = "Baggio, Marco and Ohlsson Sax, Olof and Sfondrini,
	Alessandro and Stefanski, Bogdan and Torrielli,
	Alessandro",
	title          = "{Protected string spectrum in AdS$_{3}$/CFT$_{2}$ from
		worldsheet integrability}",
	journal        = "JHEP",
	volume         = "04",
	year           = "2017",
	pages          = "091",
	doi            = "10.1007/JHEP04(2017)091",
	eprint         = "1701.03501",
	archivePrefix  = "arXiv",
	primaryClass   = "hep-th",
	reportNumber   = "NORDITA-2017-5, DMUS-MP-17-01",
	SLACcitation   = "
}
@article{Chervonyi:2016ajp,
	author         = "Chervonyi, Yuri and Lunin, Oleg",
	title          = "{Supergravity background of the $\lambda$-deformed AdS$_3
		\times$ S$^3$ supercoset}",
	journal        = "Nucl. Phys.",
	volume         = "B910",
	year           = "2016",
	pages          = "685-711",
	doi            = "10.1016/j.nuclphysb.2016.07.023",
	eprint         = "1606.00394",
	archivePrefix  = "arXiv",
	primaryClass   = "hep-th",
	SLACcitation   = "
}
@article{Wulff:2018aku,
	author         = "Wulff, Linus",
	title          = "{Trivial solutions of generalized supergravity vs
		non-abelian T-duality anomaly}",
	journal        = "Phys. Lett.",
	volume         = "B781",
	year           = "2018",
	pages          = "417-422",
	doi            = "10.1016/j.physletb.2018.04.025",
	eprint         = "1803.07391",
	archivePrefix  = "arXiv",
	primaryClass   = "hep-th",
	SLACcitation   = "
}
@article{Borsato:2018spz,
	author         = "Borsato, Riccardo and Wulff, Linus",
	title          = "{Marginal deformations of WZW models and the classical
		Yang?Baxter equation}",
	journal        = "J. Phys.",
	volume         = "A52",
	year           = "2019",
	number         = "22",
	pages          = "225401",
	doi            = "10.1088/1751-8121/ab1b9c",
	eprint         = "1812.07287",
	archivePrefix  = "arXiv",
	primaryClass   = "hep-th",
	reportNumber   = "NORDITA 2018-122",
	SLACcitation   = "
}
@article{Hoare:2016hwh,
	author         = "Hoare, Ben and van Tongeren, Stijn J.",
	title          = "{On jordanian deformations of AdS$_5$ and supergravity}",
	journal        = "J. Phys.",
	volume         = "A49",
	year           = "2016",
	number         = "43",
	pages          = "434006",
	doi            = "10.1088/1751-8113/49/43/434006",
	eprint         = "1605.03554",
	archivePrefix  = "arXiv",
	primaryClass   = "hep-th",
	reportNumber   = "HU-EP-16-15, HU-MATH-16-10",
	SLACcitation   = "
}
@article{vanTongeren:2019dlq,
	author         = "van Tongeren, Stijn J.",
	title          = "{Unimodular jordanian deformations of integrable
		superstrings}",
	year           = "2019",
	eprint         = "1904.08892",
	archivePrefix  = "arXiv",
	primaryClass   = "hep-th",
	SLACcitation   = "
}
@article{Borsato:2018spz,
	author         = "Borsato, Riccardo and Wulff, Linus",
	title          = "{Marginal deformations of WZW models and the classical
		Yang?Baxter equation}",
	journal        = "J. Phys.",
	volume         = "A52",
	year           = "2019",
	number         = "22",
	pages          = "225401",
	doi            = "10.1088/1751-8121/ab1b9c",
	eprint         = "1812.07287",
	archivePrefix  = "arXiv",
	primaryClass   = "hep-th",
	reportNumber   = "NORDITA 2018-122",
	SLACcitation   = "
}
@article{Sfetsos:2014cea,
	author         = "Sfetsos, Konstantinos and Thompson, Daniel C.",
	title          = "{Spacetimes for $\lambda$-deformations}",
	journal        = "JHEP",
	volume         = "12",
	year           = "2014",
	pages          = "164",
	doi            = "10.1007/JHEP12(2014)164",
	eprint         = "1410.1886",
	archivePrefix  = "arXiv",
	primaryClass   = "hep-th",
	SLACcitation   = "
}
@article{Hollowood:2014qma,
	author         = "Hollowood, Timothy J. and Miramontes, J. Luis and Schmidtt, David M.",
	title          = "{An integrable deformation of the $AdS_5 \times S^5$ superstring}",
	journal        = "J. Phys.",
	volume         = "A47",
	year           = "2014",
	number         = "49",
	pages          = "495402",
	doi            = "10.1088/1751-8113/47/49/495402",
	eprint         = "1409.1538",
	archivePrefix  = "arXiv",
	primaryClass   = "hep-th",
	SLACcitation   = "
}
@article{Georgiou:2018gpe,
	author         = "Georgiou, George and Sfetsos, Konstantinos",
	title          = "{The most general $\lambda$-deformation of CFTs and
		integrability}",
	journal        = "JHEP",
	volume         = "03",
	year           = "2019",
	pages          = "094",
	doi            = "10.1007/JHEP03(2019)094",
	eprint         = "1812.04033",
	archivePrefix  = "arXiv",
	primaryClass   = "hep-th",
	SLACcitation   = "
}
@article{Kawaguchi:2014qwa,
	author         = "Kawaguchi, Io and Matsumoto, Takuya and Yoshida,
	Kentaroh",
	title          = "{Jordanian deformations of the $AdS_5 x S^5$
		superstring}",
	journal        = "JHEP",
	volume         = "04",
	year           = "2014",
	pages          = "153",
	doi            = "10.1007/JHEP04(2014)153",
	eprint         = "1401.4855",
	archivePrefix  = "arXiv",
	primaryClass   = "hep-th",
	reportNumber   = "KUNS-2477, ITP-UU-14-05, SPIN-14-05",
	SLACcitation   = "
}
@article{Orlando:2010yh,
	author         = "Orlando, Domenico and Reffert, Susanne and Uruchurtu,
	Linda I.",
	title          = "{Classical Integrability of the Squashed Three-sphere,
		Warped AdS3 and Schroedinger Spacetime via T-Duality}",
	journal        = "J. Phys.",
	volume         = "A44",
	year           = "2011",
	pages          = "115401",
	doi            = "10.1088/1751-8113/44/11/115401",
	eprint         = "1011.1771",
	archivePrefix  = "arXiv",
	primaryClass   = "hep-th",
	reportNumber   = "IPMU10-0195, IMPERIAL-TP-2010-LIU-02",
	SLACcitation   = "
}
@article{OhlssonSax:2011ms,
	author         = "Ohlsson Sax, Olof and Stefanski, Jr., B.",
	title          = "{Integrability, spin-chains and the AdS3/CFT2
		correspondence}",
	journal        = "JHEP",
	volume         = "08",
	year           = "2011",
	pages          = "029",
	doi            = "10.1007/JHEP08(2011)029",
	eprint         = "1106.2558",
	archivePrefix  = "arXiv",
	primaryClass   = "hep-th",
	reportNumber   = "UUITP-17-11",
	SLACcitation   = "
}
@article{Sundin:2012gc,
	author         = "Sundin, Per and Wulff, Linus",
	title          = "{Classical integrability and quantum aspects of the
		AdS(3) x S(3) x S(3) x S(1) superstring}",
	journal        = "JHEP",
	volume         = "10",
	year           = "2012",
	pages          = "109",
	doi            = "10.1007/JHEP10(2012)109",
	eprint         = "1207.5531",
	archivePrefix  = "arXiv",
	primaryClass   = "hep-th",
	reportNumber   = "MIFPA-12-26",
	SLACcitation   = "
}
@article{Borsato:2014hja,
	author         = "Borsato, Riccardo and Ohlsson Sax, Olof and Sfondrini,
	Alessandro and Stefanski, Bogdan",
	title          = "{The complete AdS$_{3} \times$ S$^3 \times$ T$^4$
		worldsheet S matrix}",
	journal        = "JHEP",
	volume         = "10",
	year           = "2014",
	pages          = "66",
	doi            = "10.1007/JHEP10(2014)066",
	eprint         = "1406.0453",
	archivePrefix  = "arXiv",
	primaryClass   = "hep-th",
	reportNumber   = "IMPERIAL-TP-OOS-2014-03, HU-MATHEMATIK-2014-11,
	HU-EP-14-19, SPIN-14-15, ITP-UU-14-17",
	SLACcitation   = "
}
\end{bibtex}

\bibliographystyle{nb}
\bibliography{\jobname}

\end{document}